# de Haas–van Alphen spectroscopy and fractional quantization of magnetic-breakdown orbits in moiré graphene

Matan Bocarsly[1]†, Matan Uzan[1]†, Indranil Roy[1]†, Sameer Grover[1], Jiewen Xiao[1], Zhiyu Dong[2], Mikhail Labendik[1], Aviram Uri[2], Martin E. Huber[3], Yuri Myasoedov[1], Kenji Watanabe[4], Takashi Taniguchi[5], Binghai Yan[1], Leonid S. Levitov[2]#, and Eli Zeldov[1]*

[1]Department of Condensed Matter Physics, Weizmann Institute of Science; Rehovot 7610001, Israel

[2]Department of Physics, Massachusetts Institute of Technology; Cambridge, Massachusetts 02139, USA

[3]Departments of Physics and Electrical Engineering, University of Colorado Denver; Denver, Colorado 80217, USA

[4]Research Center for Functional Materials, National Institute for Materials Science; 1-1 Namiki, Tsukuba 305-0044, Japan

[5]International Center for Materials Nanoarchitectonics, National Institute for Materials Science; 1-1 Namiki, Tsukuba 305-0044, Japan

†These authors contributed equally to this work

#Email: levitov@mit.edu

*Corresponding author. Email: eli.zeldov@weizmann.ac.il

Quantum oscillations originating from the quantization of the electron cyclotron orbits provide ultrasensitive diagnostics of electron bands and interactions in novel materials. We report on the first direct-space nanoscale imaging of the thermodynamic magnetization oscillations due to the de Haas-van Alphen effect in moiré graphene. Scanning by SQUID-on-tip in Bernal bilayer graphene crystal-axis-aligned to hBN reveals abnormally large magnetization oscillations with amplitudes reaching 500 $\mu_B$ /electron in weak magnetic fields, unexpectedly low frequencies, and high sensitivity to the superlattice filling fraction. The oscillations allow us to reconstruct the complex band structure in exquisite detail, revealing narrow moiré bands with multiple overlapping Fermi surfaces separated by unusually small momentum gaps. We identify distinct sets of oscillations that violate the textbook Onsager Fermi surface sum rule, signaling formation of exotic broad-band particle-hole superposition states induced by coherent magnetic breakdown.

Oscillations in the thermodynamic and transport properties of metals subject to an external magnetic field are a fundamental quantum effect originating from the quantization of the cyclotron orbit areas. In 2D systems, the periodicity of quantum oscillations (QOs), explained by discrete Landau levels (LLs), is related in a universal way to the applied field strength and the Fermi surface (FS) geometry. The oscillations carry a wealth of information about the FS and are indispensable for resolving the band structure of moiré materials, where the presence of a superlattice potential and enhanced electron-electron interactions lead to formation of narrow minibands with multiple FSs and symmetry broken states (1–7). Quantum oscillations (QOs) can also reveal the band topology (8, 9) and strain-induced pseudomagnetic fields in graphene (10–12).

In bulk materials, QOs can be detected by measuring magnetization oscillations due to the de Haas-van Alphen (dHvA) effect. These thermodynamic oscillations, however, are usually experimentally inaccessible in 2D electron systems since the signal scales with the sample volume and is therefore extremely weak in 2D. This limits the studies of 2D electron systems mostly to the non-thermodynamic Shubnikov-de Haas (SdH) oscillations in transport coefficients. Nevertheless, several studies have succeeded in resolving magnetization oscillations in 2D electron gas (2DEG) in GaAs heterostructures using mm size samples (13–16), as well as in magnetically doped ZnSe (17). (17). In contrast, exfoliated clean van der Waals structures are of typical sizes limited to tens of μm, which makes observation of dHvA effect in atomically thin systems extremely challenging. Furthermore, all previous dHvA studies in 2DEG and in bulk materials have been global, providing no spatial information about the local band structure and thermodynamic electronic properties.

Here we report on the first spatial mapping of the dHvA effect in a van der Waals structure with resolution as high as 170 nm. We observe very large magnetization oscillations in moiré flat bands in Bernal-stacked bilayer graphene (BLG) aligned to hBN (2, 18, 19). The oscillations appear at low magnetic fields and at carrier densities of a few electrons per superlattice unit cell. In the integer quantum Hall effect (QHE), the periodicity of the QOs is tied in a universal manner to the carrier density through the LL degeneracy. To the contrary, here the observed oscillations display characteristic frequencies that are an order of magnitude lower and form complex spectra, revealing the coexistence of multiple FSs and allowing accurate moiré band structure reconstruction.

When a number of FSs coexist, the QOs display several frequencies, reflecting the relative size of the different Fermi pockets encircled by the cyclotron orbits (17, 20–24). In addition to these fundamental orbits, exotic electron orbits delocalized in $k$-space and supporting coherently entangled states in different bands can arise due to interpocket tunneling. Such tunneling, which is hindered by momentum conservation at zero magnetic field, is made possible at elevated fields through the coherent magnetic breakdown (CMB) mechanism (25–32). CMB has been predicted to occur in moiré graphene (23, 33), yet so far it has evaded detection. We provide the first observation of CMB in atomic van der Waals structures evidenced by uniquely rich sets of QOs. Conventional CMB is expected to occur at high fields in the vicinity of saddle points or Lifshitz transitions, across which the topology of the FS changes (23, 33, 34). To the contrary, we find a broad-band breakdown at very low applied fields, extending over several moiré bands and spanning a wide range of energies. The observed QOs indicate the occurrence of particle-hole superposition states shared by closely-proximitized bands and exhibiting a high degree of interband phase coherence. The hallmark of such states is unusual CMB oscillation frequencies violating Onsager's FS area sum rule. Instead, the QO frequencies are described by fractional Onsager quantization relations.

**Transport measurements**

Transport measurements of $\rho_{xx}$ and $\rho_{yx}$ (Figs. 1A,B) of the BLG sample (Fig. 2A) were performed at a temperature $T = 300$ mK as a function of applied out-of-plane magnetic field $B_a$ and carrier density $n$. Peaks in $\rho_{xx}$ (Fig. 1C) at $n = 4n_0 \cong \pm 3.48 \times 10^{12}$ cm$^{-2}$ indicate that the BLG is aligned to the hBN substrate with a twist of $\theta \cong 0.70°$, forming a moiré superlattice with unit cell size $\lambda \cong 11.5$ nm (*35*), where $n_0$ corresponds to one electron per moiré unit cell. The weak $\rho_{xx}$ peaks at filling factor $\nu = n/n_0 = \pm 4$ reflect minima in the density of states (DOS) and the absence of a full gap between the flat and remote bands.

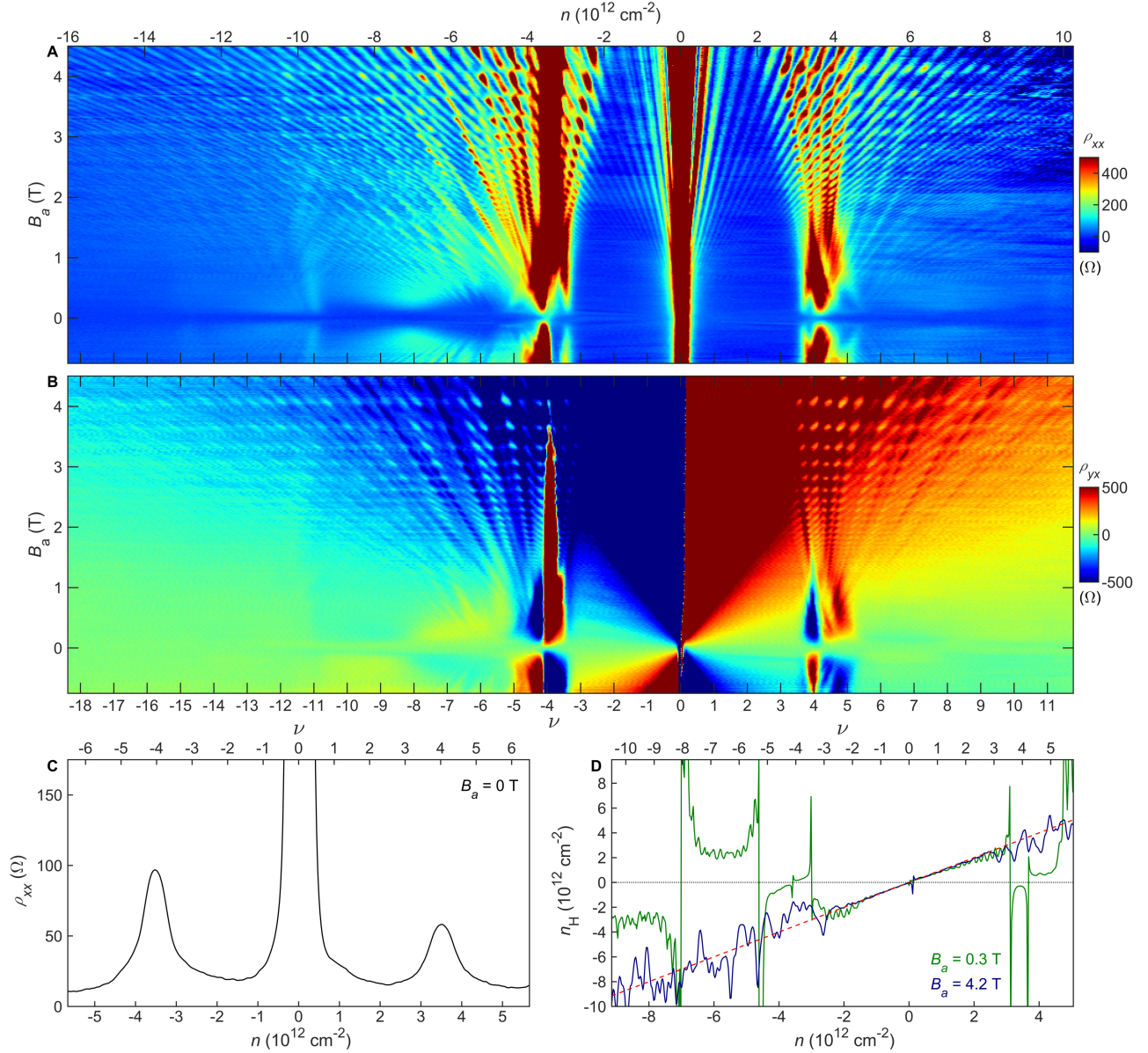

**Fig. 1. Transport measurements in BLG aligned to hBN.** (**A**) $\rho_{xx}$ vs. carrier density $n$ and magnetic field $B_a$ at $T = 300$ mK. The oscillations at high $B_a$ display four-fold degenerate LLs and Hofstadter's butterfly (Fig. S1). Resistivity values above 500 Ω are saturated for clarity. (**B**) $\rho_{yx}$ vs. $\nu$ and $B_a$. Resistivity values above 500 Ω are saturated. (**C**) $\rho_{xx}$ vs. $\nu$ at $B_a = 0$ T. (**D**) Hall carrier density $n_H = B_a/(e\rho_{yx})$ vs. $n$ derived from $\rho_{yx}$ at $B_a = 0.3$ T (green) and $B_a = 4.2$ T (blue). The dashed red line shows a slope of 1 corresponding to $n_H = n$.

At low fields, $\rho_{yx}$ shows several sign reversals in Fig. 1B. The Hall carrier density $n_H$ derived at $B_a = 300$ mT (Fig. 1D, green line) reveals a van Hove singularity at $\nu \cong 3.5$, accompanied by a change in carrier type from electrons to holes. A similar behavior is found at $\nu \cong -3.5$. Several additional $n_H$ sign reversals are observed at higher $|\nu|$, consistent with the presence of several remote bands. Notably, at higher fields, the sign reversals disappear, as shown by the blue line in Fig. 1D (taken at $B_a = 4.2$ T), and $n_H$ shows a continuous evolution with doping following $n_H = n$ (dashed red line), a behavior characteristic of a single band.

At elevated $B_a$ a Landau fan originating from CNP is visible at all fillings (Fig. 1A), corresponding to four-fold degenerate LLs due to spin and valley degeneracies in graphene (Fig. S1B). Additionally, Hofstadter patterns are visible as horizontal lines periodic with $\phi_0/B_a$, arising from the interference of the moiré unit cell with the area occupied by a flux quantum $\phi_0 = h/e$ (where $h$ is Planck constant and $e$ is the elementary charge), as reported previously (*1–3*, *18*, *36*, *37*). More complicated Landau fans that originate from the vicinity of $\nu = \pm 4$ are discerned at intermediate fields. In the following, we describe local studies at fields below 350 mT, where QOs in transport measurements are hardly resolved.

**Imaging quantum oscillations**

To study the magnetization oscillations we utilize a scanning superconducting quantum interference device fabricated on the apex of a sharp pipette (SQUID-on-tip, SOT) (*38*). An Indium SOT (*39*) of about 150 nm diameter is scanned at a height of $h \approx 200$ nm above the sample surface (Fig. 2A) at $T = 300$ mK (*35*). The *dc* voltages $V_{tg}^{dc}$ and $V_{bg}^{dc}$ applied to the top and bottom Pt gates are used to control $n$. A small *ac* voltage $V_{bg}^{ac}$ of 5 to 20 mV rms is applied to the backgate, modulating the carrier density by $n^{ac}$ and the corresponding $\nu^{ac}$ by 0.004 to 0.016, and the resulting $B_z^{ac}(x, y) = n^{ac}(dB_z/dn)$ is imaged across the sample. This signal reflects the induced *ac* modulation in the local magnetization, $m_z(x, y) = dM_z(x, y)/dn$, which can be reconstructed directly from the measured $B_z^{ac}(x, y)$ by a numerical inversion (*40*) ($M_z$ is the magnetization per unit area and $m_z$ is the magnetization per excess electron; both dominated by the orbital effects (*35*)). Figure 2B shows an example of the resulting map of the local differential magnetization at $\nu = -7.65$ and $B_a = 334$ mT, displaying extremely large values of $m_z(x, y)$ reaching $\pm 500$ $\mu_B$/electron and forming patches of positive and negative $m_z$ with a characteristic size of about 1 μm. Upon varying $\nu$, the patches move across the sample and $m_z(x, y)$ reveals remarkable quasi-periodic oscillations as shown in Fig. 2C and in Movie S1. The period of the QOs and the $m_z$ amplitude vary significantly with position as seen in Figs. 2D,E. We observe these oscillations starting from applied field as low as 116 mT and up to our highest $B_a = 334$ mT at which our SOT has sufficient sensitivity (Fig. S2). Similar behavior is found at various values of displacement field $D$ (Fig. S3) and in two additional samples (Figs. S3, S4). This is the first spatially resolved measurement of the dHvA effect in graphene devices.

To investigate the origin of the thermodynamic QOs, we measure the evolution of $B_z^{ac}(x)$ with carrier density over an extended range of $\nu$ (Fig. 3A) by repeated scanning along the black dotted line in Fig. 2A, while incrementing $\nu$ in 0.006 steps at $B_a = 300$ mT. For $|\nu| \lesssim 3.5$ weak periodic oscillations in $B_z^{ac}(x)$ are discerned as shown in Fig. 3C. For $|\nu| \gtrsim 3.5$ the behavior is markedly different, characterized by the appearance of strong oscillations with significantly larger and variable periods $\Delta n$, as visualized in Figs. 3B,D. To quantify the oscillations' periodicity, we perform a fast Fourier transform (FFT) of $B_z^{ac}(\nu)$ over a narrow window of $\delta \nu = 1.11$ around a given $\nu$ (*35*). Figure 3E shows the FFT at the $x$ position marked by the white dashed line in Fig. 3A (see Movie S2 for FFT at other $x$ positions).

In the integer QHE, the frequency of the QOs as a function of $n$ is given by $f = \frac{1}{N}\frac{\phi_0}{B_a}$, where $N$ is the spin-valley degeneracy. For a given $N$, $f$ is determined solely by $B_a$ and should thus be independent of the position and moiré band filling $\nu$. For $|\nu| \lesssim 3.5$ the FFT reveals a peak at $f \cong \frac{\phi_0}{4B_a}$ which is rather independent of $\nu$ (see Movie S2 and Fig. S4 for an additional sample). This shows that the QOs at these fillings originate from the standard QHE with $N = 4$ spin and valley degenerate LLs.

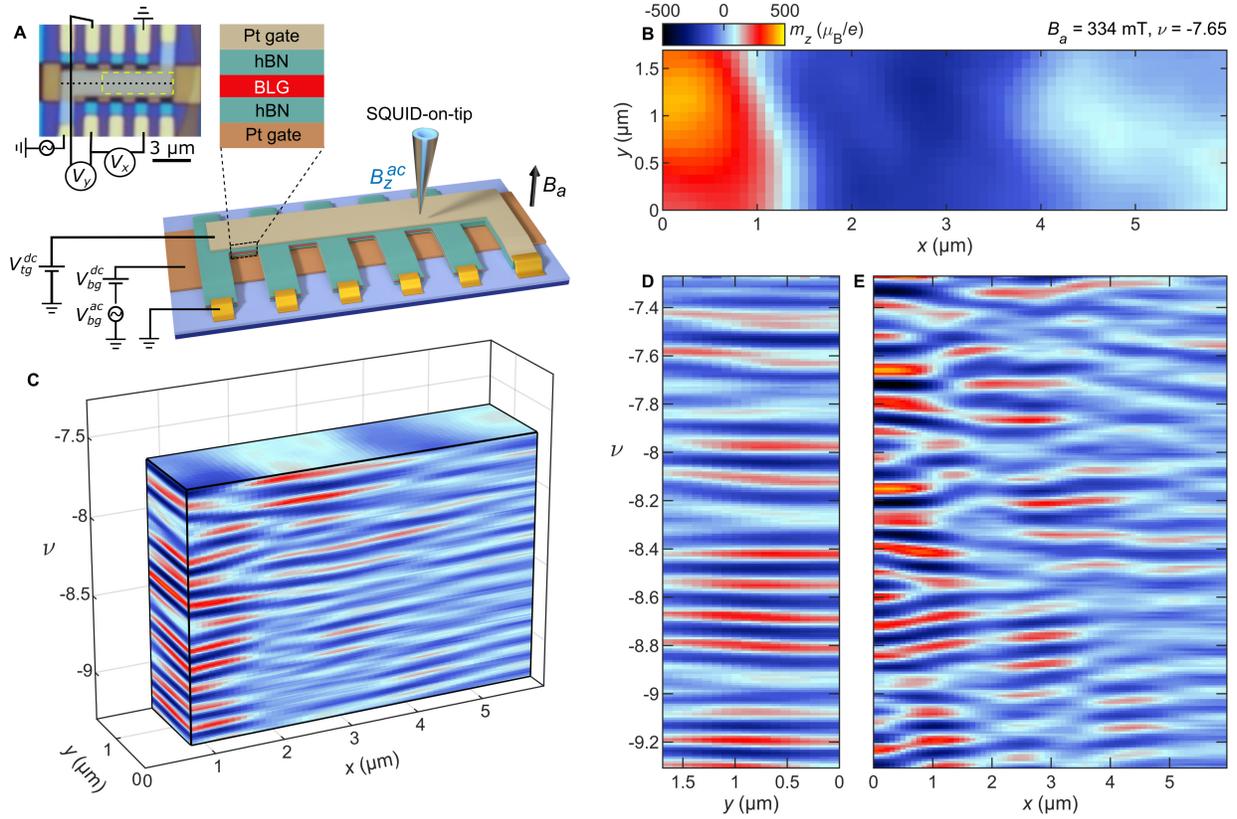

**Fig. 2. Imaging the dHvA effect.** (**A**) Top: Optical image of the BLG/hBN sample with indicated contacts for $\rho_{xx}$ and $\rho_{yx}$ measurements. The dashed yellow rectangle marks the area imaged in (B) and the dotted black line marks the line cut presented in Fig. 3. Bottom: Schematic sample structure indicating the applied top-gate and back-gate voltages, $V_{tg}^{dc}$ and $V_{bg}^{dc} + V_{bg}^{ac}$, and the corresponding $ac$ magnetic field $B_z^{ac}$ imaged by the scanning SOT. (**B**) Map of orbital magnetization at $B_a = 334$ mT, $T = 300$ mK, and $\nu = -7.65$ showing domains of positive and negative local magnetization $m_z(x, y)$ with amplitude of up to ±500 $\mu_B$/electron. The color bar applies to all the panels. (**C**) Tomographic rendering of $m_z(x, y, \nu)$ (see Movie S1). (**D**) Slice of the tomographic data $m_z(y, \nu)$ at $x = 1.36$ μm. (**E**) $m_z(x, \nu)$ slice along $y = 1.22$ μm.

At $|\nu| \gtrsim 3.5$ the QOs show a remarkably rich behavior (Fig. 3E) that departs from the standard QHE in a number of ways: (i) The frequency of the oscillations is up to one order of magnitude lower than $\frac{\phi_0}{4B_a}$. (ii) Rather than being restricted to integer fractions, $f$ varies continuously as a function of $\nu$. (iii) At higher filling factors more than one characteristic frequency is present simultaneously. (iv) $f$ varies in space as can be appreciated from Fig. 3A and Movie S2. These features reveal the presence of narrow moiré bands with overlapping FSs as discussed next.

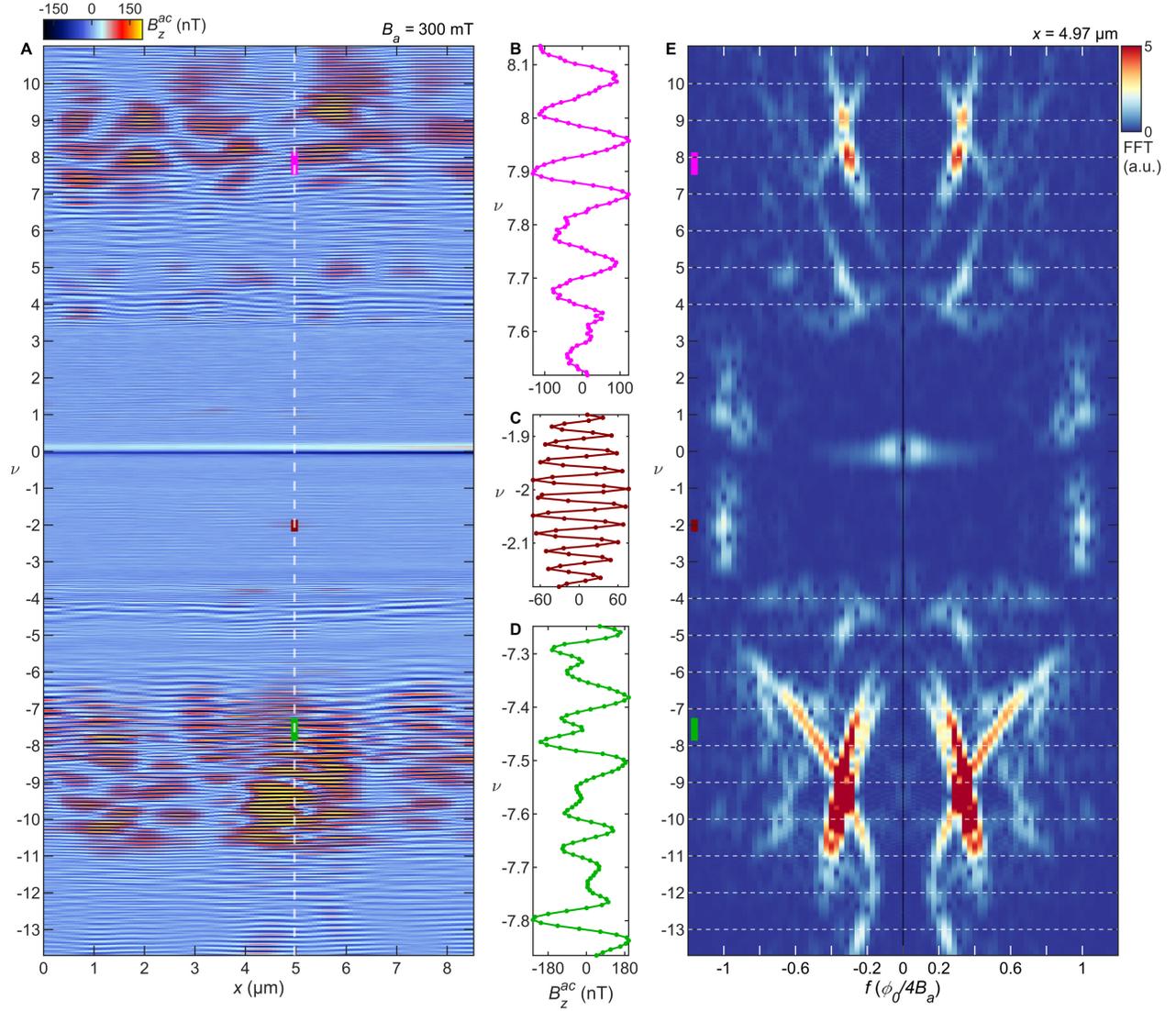

**Fig. 3. Evolution of the dHvA quantum oscillations with $\nu$ and position.** (**A**) $B_z^{ac}(x, \nu)$ measured along the dotted black line in Fig. 1A at $B_a = 300$ mT (see Fig. S2 for additional field values). At $|\nu| \lesssim 3.5$ the orbital magnetization and the corresponding $B_z^{ac}$ are weak, while for $|\nu| \gtrsim 3.5$ large $B_z^{ac}$ accompanied by pronounced low-frequency QOs are present. (**B**) Zoomed-in cross-section of $B_z^{ac}(\nu)$ along the magenta segment in (A), showing low-frequency QOs with period that gradually varies with $\nu$. (**C**) $B_z^{ac}(\nu)$ cross-section along the red segment in (A), showing high-frequency periodic oscillations due to conventional four-fold-degenerate LLs. (**D**) $B_z^{ac}(\nu)$ along the green segment in (A), revealing large-amplitude low-frequency oscillations comprising multiple frequencies. (**E**) FFT of $B_z^{ac}(\nu)$ at $x = 4.97$ μm marked by the white dotted line in (A) performed over a narrow window of $\delta\nu = 1.11$ around $\nu$ (see Movie S2 for FFT at different locations). The frequency is in units of $\phi_0/4B_a$ and both positive and negative FFT frequencies are shown for clarity. At $|\nu| \lesssim 3.5$ the QOs arise from conventional four-fold degenerate QHE LLs with $f = \frac{\phi_0}{4B_a}$. For $|\nu| \gtrsim 3.5$ the low-frequency oscillations are governed by multiple overlapping Fermi surfaces.

**Band structure of moiré bilayer graphene**

To gain further insight we perform continuum-model single-particle band structure (BS) calculations of BLG/hBN moiré (*35*, *41*). The mismatch of graphene and hBN's lattice constants creates a real space superlattice, and causes band folding into the moiré mini-Brillouin zone (mBz) (*1–3*, *41–45*). Figure 4A shows the calculated conduction C1 and valence V1 flat bands along with highly-overlapping remote valence bands V2 and V3. Following (*41*), we center the mBz around the original graphene $K$ point, and label the original hBN Brillouin zone corner $Y$ and its time reversal counterpart $X$. Generally, the inequivalent lattice sites in the hBN substrate break inversion symmetry and open a gap at the charge neutrality point (CNP). Remote bands, such as V2 and V3, however, are highly overlapping creating complex Fermi pockets. Depending on the specifics of band structure parameters, either a full gap or a minimum in DOS occurs between V1 and V2 at $\nu = -4$ (*35*).

For $|\nu| \lesssim 3.5$ the FS topology is simple with a single Fermi pocket (FP) around the $K$ point (Fig. 4B). At $|\nu| \cong 3.5$ a Lifshitz transition occurs (consistent with the observed van Hove singularity at low $B_a$ in Fig. 1D), resulting in the formation of two FPs centered around $X$ and $Y$ points (see Movie S3). With increasing $|\nu|$, the FS topology becomes more complicated, consisting of three or more FPs due to overlapping bands (Fig. 4C). Each FP accommodates independent LLs leading to QOs with multiple fundamental frequencies. Traditionally, in bulk materials, QOs are measured vs. $1/B_a$, in which case each FP contributes an oscillation with a frequency proportional to the FP area $S_i$ (*20*, *22*). This conjecture, however, does not hold for QOs measured vs. $n$, in which case the oscillation frequencies are given by $f_i = \frac{\mathcal{N}_i(\epsilon_F)}{\mathcal{N}(\epsilon_F)} \frac{\phi_0}{4B_a}$, where $\mathcal{N}_i(\epsilon_F) = \frac{1}{4\pi^2} \frac{\partial S_i}{\partial \epsilon}$ is the DOS of pocket $i$, $\mathcal{N}(\epsilon_F) = \sum_i \mathcal{N}_i(\epsilon_F)$ is the total DOS, $\epsilon_F$ is the Fermi energy, and we consider four-fold degenerate bands (*35*). This stems from the fact that varying $B_a$ affects the cyclotron motion of the entire Fermi sea, whereas varying $n$ affects the behavior only near $\epsilon_F$. Hence, upon increasing $\epsilon_F$ by $\Delta\epsilon$, one LL is added to a pocket when its $S_i$ has increased by $\Delta S_i = 4\pi^2 \mathcal{N}_i(\epsilon_F)\Delta\epsilon = \frac{4B_a}{\phi_0}$. This leads to the Onsager sum rule of the fundamental frequencies, $\sum_i f_i = f_0 = \frac{\phi_0}{4B_a}$.

Figure 4D shows the experimental FFT data from Fig. 3E overlaid with color-coded lines indicating the $f_i(\nu)$ of the different pockets calculated from the BS. For hole doping at $0 > \nu > -3.5$, only one FP around $K$ point, V1$_K$, is present, resulting in a single frequency $f_{V1K} = f_0$ (green in Fig. 4D). At the Lifshitz transition at $\nu \cong -3.5$ the FS breaks into two pockets around the mBz corners, V1$_X$ and V1$_Y$ (Movie S3). As a result, two QO frequencies coexist for a small region of $\nu$, $f_{V1X}$ and $f_{V1Y}$, until V1 (green) overlaps V2 (light brown). At $\nu < -5$, V3 band (pink) starts to be occupied forming two FPs, V3$_X$ and V3$_Y$, with increasing DOS with $|\nu|$ that coexist with the V2$_K$ FP with decreasing DOS (Fig. 4C). As a result, the two V3 frequencies, $f_{V3X}$ and $f_{V3Y}$, grow with $|\nu|$ (pink in Fig. 4D), while the $f_{V2K}$ frequency decreases (light brown). The calculated behavior for $-5 > \nu > -10.5$ closely follows the experimentally derived frequencies and their evolution with $\nu$. Similar behavior is observed for electron doping at $\nu > 3.5$.

With the above insight we note that our high sensitivity to the oscillation frequencies at fillings where multiple Fermi pockets coexist, makes the nanoscale magnetization imaging a uniquely sensitive tool for mapping the local band structure and extracting BS parameters. There has been little theoretical work and limited experimental determination of the coupling strengths between hBN and graphene. In Ref. (*41*) the tunneling strengths between overlapping boron and carbon atoms ($t_{BC}$) and between overlapping nitrogen and carbon ($t_{NC}$) are taken to be equal, or $r_{NB} \equiv t_{NC}/t_{BC} = 1$. In Ref. (*45*), $r_{NB} \cong 0.67$, based on *ab initio* DFT calculations. We find that both these parameter sets fail to fit our experimental data (Fig. S5).

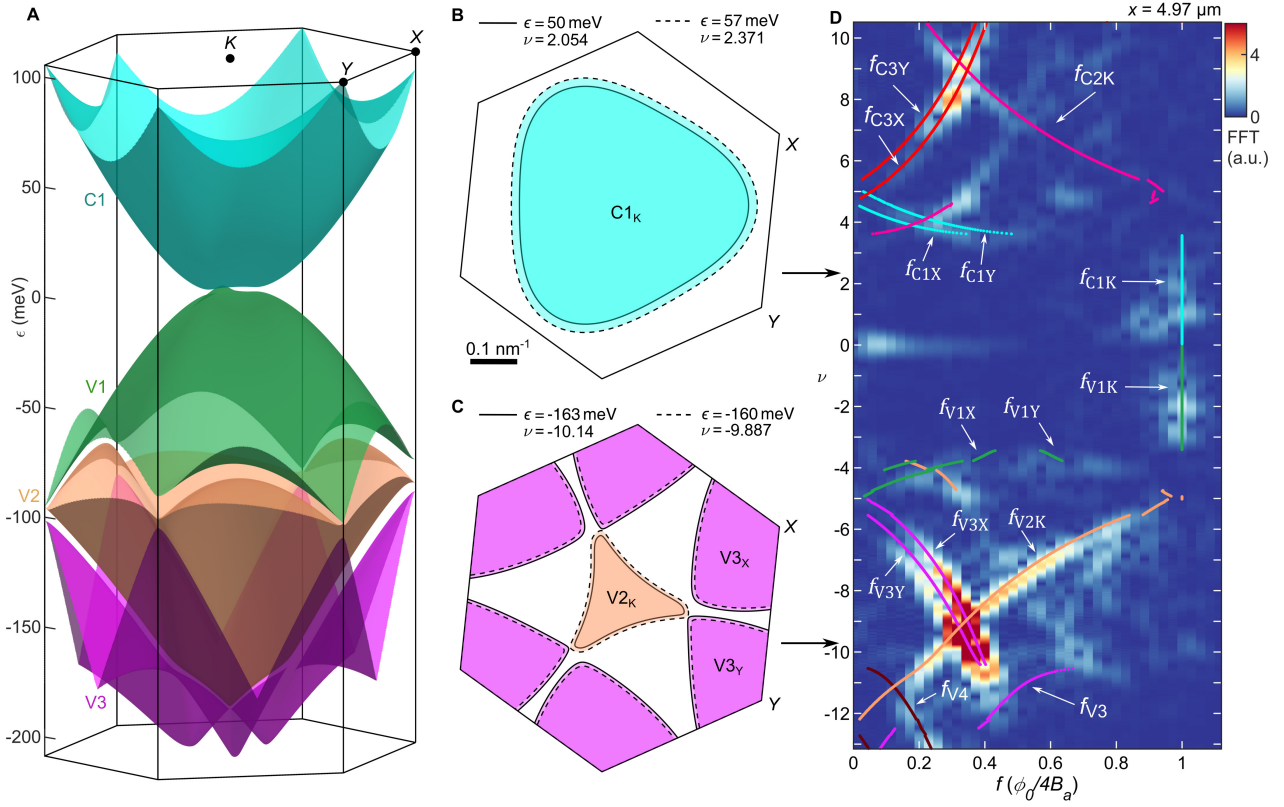

**Fig. 4. Calculation of BLG/hBN band structure and of QOs.** (**A**) Single-particle band structure calculation for a single valley ($K$) showing the conduction flat band (C1, top), valence flat band (V1, middle) and two partially overlapping remote valence bands (V2 and V3, bottom) in the moiré mini-Brillouin zone of BLG aligned to hBN with $\theta = 0.75°$. The bands are four-fold degenerate with $K'$ bands rotated by 180°. Tight-binding parameters were chosen to best fit (D) with $r_{NB} = 0.5$ and $w = 0.5$ ((*35*) and Fig. S5). (**B-C**) Example of simple (B) and complex (C) Fermi surfaces (solid contours) at $\nu = 2.054$ and $-10.14$ respectively. The dashed contours indicate the change in the areas of the Fermi pockets with small increase in $\nu$, reflecting the DOS and the QO frequencies of each pocket. (**D**) The FFT of QOs from Fig. 3E, overlaid with fundamental frequencies (lines coded by band colors) calculated by the relative DOS of each pocket, $f_i(\nu) = \frac{\mathcal{N}_i(\epsilon_F)}{\mathcal{N}(\epsilon_F)} \frac{\phi_0}{4B_a}$.

Recent STM experiments have revealed significant lattice relaxation in magic-angle twisted bilayer graphene (*46*) arising from the higher energy of *AA* stacking atomic configuration in comparison to *AB* configuration. This lattice relaxation has significant impact on the calculated BS, including gap opening between the flat and dispersive bands in magic-angle graphene (*47*, *48*), as observed experimentally (*4*, *49*). In the continuum model, this lattice relaxation is captured with the phenomenological parameter $w = t_{AA}/t_{AB} \approx 0.8$ (*50*). Recently, molecular dynamics simulations on aligned BLG/hBN heterostructures have also shown significant lattice relaxation, which has been proposed to have an effect on the topology of the system (*19*).

By fitting the high resolution QOs we find $r_{NB} \cong 0.5$ and $w \cong 0.5$, significantly lower than evaluated previously, leading to larger band overlaps with smaller energy gaps (Fig. S5) and to magnetic breakdown as discussed next. This finding of strong lattice relaxation in aligned BLG/hBN calls for further exploration of lattice relaxation mechanisms and its effect on hBN aligned moiré heterostructures.

**Magnetic breakdown**

The BS calculations with enhanced lattice relaxation provide a good description of the observed fundamental QO frequencies. Yet, there are a number of prominent lines in Fig. 4D that are not accounted for by the calculated $f_i(\nu)$. Moreover, these lines do not obey the Onsager band area sum rule $\sum_i f_i = f_0$ and cannot be explained by simple harmonics of $f_i$. These unaccounted-for lines indicate the presence of new electron orbits that encompass areas outside the closed FS contours. Such trajectories can be facilitated by interband electron tunneling due to CMB mechanism, which has been widely investigated in bulk metals (*27*), but so far has not been identified in 2D vdW materials. When two Fermi pockets are separated by a small momentum-gap $\Delta k$, the magnetic-field-induced interpocket tunneling occurs with probability

$$P \cong e^{-\frac{B_{MB}}{B_a}}, \qquad B_{MB} = \frac{\phi_0}{2}\left(\frac{\Delta k^3}{\frac{1}{R_1}+\frac{1}{R_2}}\right)^{1/2} \cong \frac{\phi_0}{2}\Delta k^2,$$

where $B_{MB}$ is the breakdown field, and $R_1$ and $R_2$ are the $k$-space radii of curvature of the two FS in the gap region (*27*). For $B_{MB} \lesssim B_a = 0.3$ T this requires $\Delta k \lesssim 0.012$ nm$^{-1}$. Our BS calculations indeed show very small gaps between Fermi pockets in remote bands (Fig. 5B).

To derive the MB orbits and their corresponding QO frequencies $f_{MB}$, we analyze two prominent unaccounted for lines in the experiment at $5 < \nu < 10.5$ and $-12.5 < \nu < -10$ (Figs. 5C,E, bright green). Figure 5A shows the FS structure at $\nu = 8.4$, displaying electron pockets at $X$ and $Y$ originating from C3 band (red), and a hole pocket at $K$ from C2 band (dark pink). The sharp touching points between the C3$_Y$ and C2$_K$ pockets are characterized by a very small momentum gap $\Delta k \cong 0.010$ nm$^{-1}$, while the gaps between the C3$_X$ and C2$_K$ pockets have an even smaller $\Delta k \cong 0.005$ nm$^{-1}$ as shown in Fig. 5B inset, leading to CMB at low fields. Moreover, in contrast to the common situations where close proximity of FPs is limited to the vicinity of Lifshitz transitions, the unique highly overlapping BS of relaxed BLG/hBN leads to small gaps extending over almost the entire energy range of the remote bands with sharp ridges that closely follow each other (Fig. 5B). The tunneling between the FPs can give rise to a number of extended equi-energy electron orbits. The green dashed line in Fig. 5A shows the shortest orbit that traces 2/3 of the circumference of the two electron pockets C3$_X$ and C3$_Y$ and 1/3 of the circumference of the hole pocket C2$_K$. As a result, the QO frequency of this orbit is given by a fractional Onsager relation,

$$f_{MB} = \frac{1}{3}(2f_{C3X} + 2f_{C3Y} + f_{C2K}),$$

where $f_i$ are the corresponding QO frequencies of the individual FPs, in contrast to common CMB behavior with integer Onsager relations (*27, 32*). Figure 5C shows a good fit between the calculated $f_{MB}$ (green) and the experimentally unaccounted for QO line. Interestingly, a gap of $\Delta k \cong 0.01$ nm$^{-1}$ corresponds to tunneling probability $P \approx 0.5$, allowing the carriers to orbit along both the closed FSs and along the CMB trajectories. As a result, both the fundamental (red and pink) and the CMB frequency lines (green) are observed concurrently in Fig. 5C.

Figure 5D shows the BS at $\nu = -12.28$, displaying three degenerate hole pockets in V4 band and an electron pocket at $K$ in V3 band. CMB creates an electron trajectory that flows along the inner and outer edges of V3 and V4 pockets (green dashed line in Fig. 5D), which explains well the unaccounted QO line in Fig. 5E as marked by the calculated green $f_{MB}$ line. Notably, the fundamental $f_{V3}$ line (purple) is essentially invisible in the experiment. This can be understood in view of the extremely small gap between the V3 and V4 pockets with $\Delta k \cong 0.002$ nm$^{-1}$ and very small radii of curvature $R$. As a result, the electrons tunnel between the pockets with probability $P \cong 1$, leaving essentially no carriers that circulate exclusively in the V3 pocket and hence no detectable $f_{V3}$. Moreover, in contrast to the usual CMB behavior (*32*) and unlike the $f_{MB}$ line in the conduction bands (green in Fig. 5C), the $f_{MB}$ line in Fig. 5E cannot be expressed as either integer or fractional Onsager combination of the fundamental frequencies due to nontrivial evolution of the FS with doping.

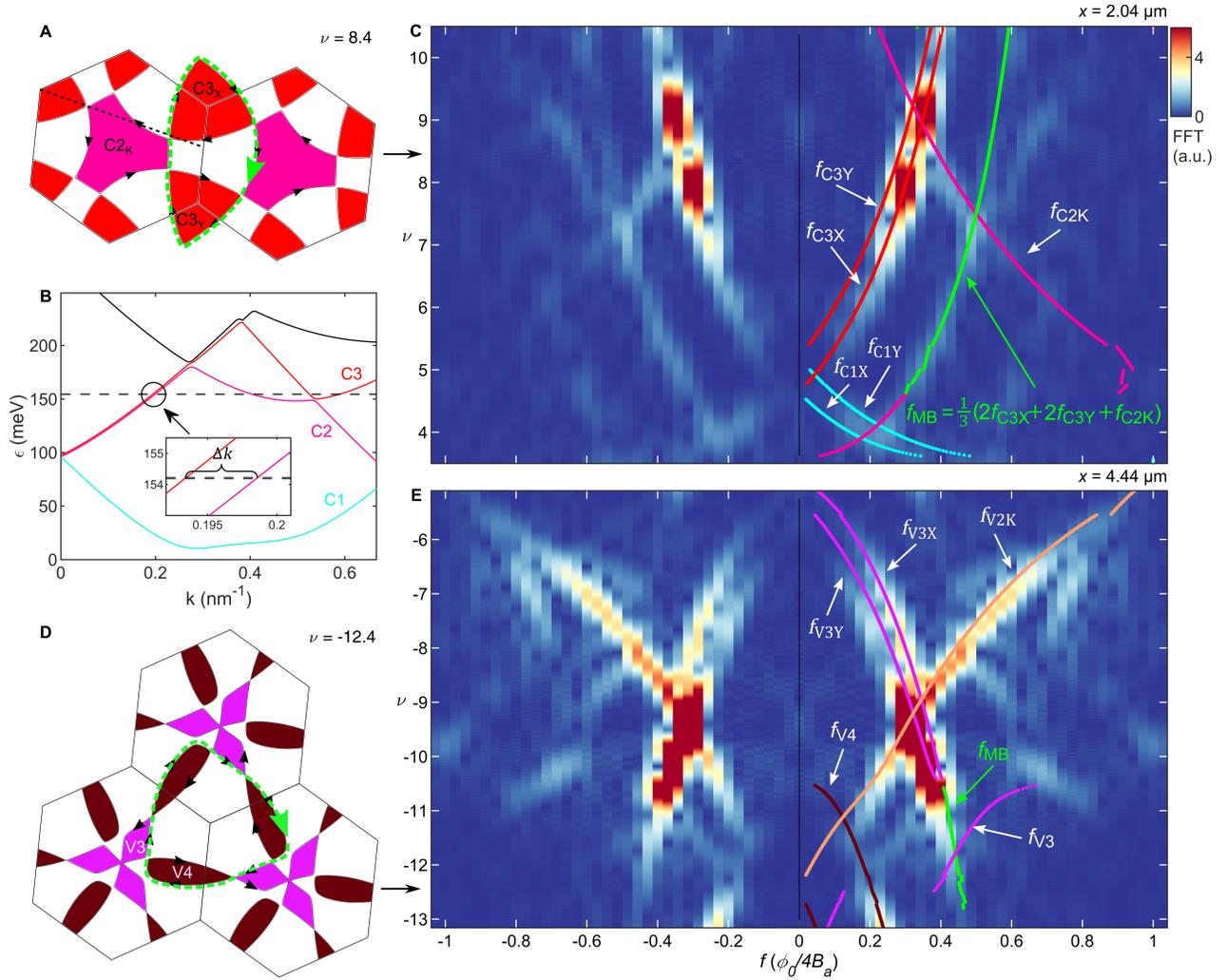

**Fig. 5. Coherent magnetic breakdown.** (**A**) Constant energy band structure cut at $\nu = 8.4$ showing the occupied $C3_X$ and $C3_Y$ electron pockets (red, clockwise black arrows) and the C2K hole pocket (dark pink, counter-clockwise arrows). The dashed green trajectory indicates the shortest magnetic breakdown orbit. (**B**) Cut of the band structure along dashed black line in (A). Dashed line at $\epsilon = 154.2$ meV corresponds to the energy value in (A). The gap between C2$_K$ and C3$_X$ pockets remains very small over a large range of energies. Inset: Zoom in showing a gap of $\Delta k \cong 0.005$ nm$^{-1}$. (**C**) Expanded view of FFT of $B_z^{ac}(\nu)$ at $x = 2.04$ μm showing a pronounced frequency line for $4.5 < \nu < 10$ that does not obey the sum rule. This frequency is accounted for by the CMB orbit in (A), resulting in $f_{MB} = \frac{1}{3}(2f_{C3X} + 2f_{C3Y} + f_{C2K})$ (green line). Positive and negative FFT frequencies are shown for clarity. (**D**) Constant energy band structure cut at $\nu = -12.28$ showing three degenerate electron pockets from V4 band (brown) and one hole pocket from V3 band (light purple). The dashed green trajectory indicates the shortest CMB orbit. (**E**) FFT of $B_z^{ac}(\nu)$ at $x = 4.44$ μm with overlaid calculated $f_{MB}$ (green line).

## Discussion

Due to the extreme sensitivity of the QOs to the band structure details, our measurement provides a unique probe of multi-band moiré FSs and their low-energy electronic properties. Crucially, the ability to detect thermodynamic QOs at low fields, allows probing of the BS with high energy resolution and without perturbing it by high magnetic fields. In particular, our results show that the hBN-graphene coupling is substantially weaker than previously estimated values, giving $r_{NB} = t_{NC}/t_{BC} \cong 0.5$ and a large lattice relaxation with $w = t_{AA}/t_{AB} \cong 0.5$, a direct demonstration of a weak moiré potential with strongly overlapping bands and small energy gaps. This fine potential modulation is readily washed out at elevated magnetic fields leading to a full breakdown in which the carriers orbit mostly along the original BLG FS unperturbed by the moiré potential (Fig. S6 and (*35*)). Indeed, the $n_H$ measured at $B_a = 4.2$ T in Fig. 1D, shows essentially no signs of the moiré multiband structure resolved at low fields.

Our findings open up a previously uncharted regime of exotic CMB physics at unusually low $B_a$, arising due to cyclotron orbits delocalized in $k$-space and supporting states coherently entangled among different sub-bands. This regime is manifested through QOs that do not obey Onsager's FS area sum rule. Instead, a fractional Onsager quantization relation is observed that indicates the occurrence of particle-hole superposition states shared by adjacent bands and exhibiting a high degree of interband phase coherence. Remarkably, the real-space cyclotron radius of the observed CMB orbits in Fig. 5A is as large as $R_c = \frac{\phi_0}{2\pi B_a}\left(\frac{S}{\pi}\right)^{1/2} \cong 500$ nm, a value comparable to the characteristic scales of disorder and sample dimensions (*35*). The particle-hole coherence induced by CMB is an appealing and not-yet-explored direction for band engineering in moiré materials.

**Acknowledgments.** The authors thank Erez Berg, Ady Stern, Igor Rozhansky, and Nurit Avraham Tayar for useful discussions.

**Funding:**

European Union (ERC, MoireMultiProbe - 101089714). "Views and opinions expressed are however those of the author(s) only and do not necessarily reflect those of the European Union or the European Research Council. Neither the European Union nor the granting authority can be held responsible for them." (EZ)

Minerva Foundation with funding from the Federal German Ministry of Education and Research Grant No. 140687 (EZ)

United States - Israel Binational Science Foundation (BSF) Grant no 2022013 (EZ, LSL)

Andre Deloro Prize for Scientific Research and Leona M. and Harry B. Helmsley Charitable Trust grant 2112-04911 (EZ)

Science and Technology Center for Integrated Quantum Materials, NSF Grant No. DMR1231319, and Army Research Office Grant W911NF-18-1-0116 (LSL)

Sagol Weizmann-MIT Bridge Program (EZ, LSL)

European Research Council (ERC Consolidator Grant "NonlinearTopo", No. 815869) (BY)

Israel Science Foundation ISF grant No. 2932/21 (BY)

JSPS KAKENHI grant numbers 19H05790, 20H00354 and 21H05233 (KW, TT)

MIT Pappalardo Fellowship (AU)

VATAT Outstanding Postdoctoral Fellowship in Quantum Science and Technology (AU)

**Author contributions:**

Local magnetization measurements: MB, IR, SG

Sample fabrication and transport measurements: MU

SOT fabrication and tuning fork feedback: IR, ML, YM

Development of experimental setup: AU

Design and building of the SOT readout system: M.E.H.

Band structure calculations: JX, BY, MB

Development of theoretical model (CMB): MB, EZ, LSL, ZD

Data analysis: MB, IR, MU, SG

hBN crystals: KW, TT

Writing of original manuscript: MB, EZ, LSL, JX, MU, IR, ZD

Editing and review of manuscript: all authors.

**Competing interests:** The authors declare no competing interests.

**Data and materials availability:** The data that support the findings of this study are available from the corresponding authors on reasonable request. The band structure calculations codes used in this study are available from the corresponding authors on reasonable request.

## Materials and Methods

### Graphene device fabrication

The hBN-encapsulated BLG heterostructures were fabricated using the dry-transfer method. The flakes were first exfoliated onto a Si/SiO$_2$ (285 nm) substrate and picked up using a polycarbonate (PC) on a polydimethylsiloxane (PDMS) dome stamp. The number of layers of the graphene flakes was determined using Raman microscopy (*51*), and the orientation of the crystallographic axes of both the hBN and BLG were identified from their straight edges. To avoid misalignment of 30°, the chirality of the edges (zigzag/armchair) was determined from the dependence of the second harmonic generation (SHG) on the polarization angle (*52, 53*) in the hBN flakes and in the ABA stacked trilayer graphene adjacent to the BLG. The Raman and SHG measurements were performed using WITec alpha300 R Raman imaging microscope using wavelengths of 532 and 1064 nm. During the dry-transfer process, the crystal axes of the hBN and BLG were aligned by a mechanical rotation stage. The stacks were then released onto a pre-annealed Ti (2 nm)/Pt (10 nm) bottom gate, patterned on Si/SiO$_2$ wafer. The doped Si substrate was also used to gate the contact regions of the device. The finalized stacks were annealed in vacuum at 500 °C for strain release (*54*). A Ti (2nm)/ Pt (10nm) top gate was then deposited on top of the stack. The 1D contacts were formed by SF$_6$ and O$_2$ plasma etching followed by evaporating Cr (4 nm)/Au (60-90 nm). Then the devices were etched into a Hall bar geometry. Finally, atomic force microscopy (AFM) scanning in contact mode was deployed to sweep off resist residues from the surface.

Device summary:

**Device 1** (presented in the main text): Ti (2nm)/Pt (10nm) top and bottom gates, the top and bottom hBN originate from the same flake with a thickness of $\cong$ 24 nm, $C_g$ = 110.4 nF·cm$^{-2}$ = 6.91×10$^{11}$ $e$ cm$^{-2}$V$^{-1}$, moiré unit cell area $a_m \cong$ 114.8 nm$^2$, moiré lattice constant $\lambda \cong$ 11.5 nm, hBN twist angle $\theta \cong$ 0.70°.

**Device 2**: Ti (2nm)/Pt (10nm) top and bottom gates, the top and bottom hBN originate from the same flake with a thickness of $\cong$ 30 nm, $C_g$ = 84.5 nF·cm$^{-2}$ = 5.28×10$^{11}$ $e$ cm$^{-2}$V$^{-1}$, moiré unit cell area $a_m \cong$ 71.2 nm$^2$, moiré lattice constant $\lambda \cong$ 9.1 nm, and bottom hBN twist angle $\theta \cong$ 1.19°.

**Device 3**: Ti (2nm)/Pt (10nm) top and bottom gates, the top and bottom hBN originate from the same flake with a thickness of $\cong$ 30 nm, $C_g$ = 84.5 nF·cm$^{-2}$ = 5.28×10$^{11}$ $e$ cm$^{-2}$V$^{-1}$, moiré unit cell area $a_m \cong$ 60.2 nm$^2$, moiré lattice constant $\lambda \cong$ 8.3 nm, hBN twist angle $\theta \cong$ 1.37°.

### SOT fabrication and imaging of local magnetization

Indium SOTs of about 150 nm diameter were fabricated as described previously (*38, 39, 55*). The magnetic imaging measurements were performed at $T$ = 300 mK using a cryogenic SQUID series array amplifier (SSAA) (*56, 57*). The SOT had a magnetic field sensitivity of down to 8 nT/Hz$^{1/2}$ at frequencies above a few kHz in the field range of 116 to 334 mT. For height control, the SOT was attached to a quartz tuning fork as described in (*58*), which was electrically excited at its resonance frequency of about 33 kHz. The $B_z^{ac}$ images were acquired with pixel size of about 85 nm and acquisition time of 0.3 to 1 s/pixel. The 2D $B_z^{ac}(x, y)$ images were used to reconstruct the magnetization $m_z(x, y)$ using numerical inversion procedure described in (*40*) as shown in Fig. 2. Since the reconstruction of $m_z$ requires 2D $B_z^{ac}(x, y)$ information, the acquired 1D line scans showing the quantum oscillations vs. $\nu$ in Fig. 3 are presented as the raw data of $B_z^{ac}(x)$.

The measured signal $B_z^{ac} = n^{ac}(dB_z/dn)$ is proportional to the modulation in the carrier density $n^{ac}$ induced by $V_{bg}^{ac}$ voltage applied to the backgate at a frequency of $f \cong$ 6 kHz. It is therefore desirable to use large $n^{ac}$ to improve the signal to noise ratio. In presence of QOs, however, $n^{ac}$ has to be substantially smaller than the period of the oscillations, $\Delta n$. We have therefore optimized $V_{bg}^{ac}$ for each set of measurements. Note however, that in the case of multiple oscillation frequencies, optimization of $V_{bg}^{ac}$ was done for the lower frequencies resulting in partial suppression of the measured amplitudes of the higher frequency oscillations. The FFTs of the local QOs were performed using a Hamming window of full width $\delta\nu$ around a given $\nu$.

## Supplementary Text

### Transport measurements

Four-point transport measurements were performed at $T = 300$ mK using low-noise voltage amplifiers and standard lock-in techniques with an $ac$ bias current of $I^{ac} = 10$ nA rms at about 11 Hz. Figures S1A,C show the longitudinal resistivity $\rho_{xx}$ and the Hall conductivity $\sigma_{xy}$ vs. carrier density $n$ and $B_a$ for Device 1. The four-fold degenerate Landau fan extending from the CNP is traced in Fig. S1B. By fitting this fan, we extract the gate capacitance, $C_g = 110.4$ nF·cm$^{-2}$ = $6.91{\times}10^{11}$ $e$ cm$^{-2}$V$^{-1}$. The carrier density corresponding to four carriers per moiré cell, $\pm 4n_0$, is extracted from the position of the peak in $\rho_{xx}$ on the hole side (Fig. 1C), as BS calculations show the hole peak to be closer to the band edge. The low resistance of the satellite peaks are consistent with minima in DOS without opening of a full gap between the flat and remote bands. The moiré unit cell area $a_m$, the lattice constant $\lambda$, and the hBN rotation angle $\theta$ are calculated from $n_0$ using the following relations:

$$a_m = \frac{1}{n_0}, \qquad \lambda = \sqrt{\frac{2a_m}{\sqrt{3}}} = \frac{(1+\delta)a_0}{\sqrt{2(1+\delta)(1-\cos\theta)+\delta^2}},$$

where $a_0 = 0.246$ nm is the graphene lattice constant and $\delta = \frac{a_{BN}}{a_0} - 1 = 0.018$ is the lattice mismatch between graphene and hBN (*42*).

### Magnetic field dependence of quantum oscillations

Since the SOT sensitivity is periodic with magnetic field (*38*), the local imaging of the QOs is performed only at specific values of $B_a$ at which our SOT sensitivity is the highest. In addition to the 334 and 300 mT data presented in Figs. 1 and 2, Fig. S2 shows quantum oscillations at $B_a = 116$, 209, and 244 mT in the high hole-doping region $\nu \lesssim -3.5$ (left column). Line cuts through the data at position $x$ marked by the white dashed lines are shown in the central column, while the FFT of the line cuts is presented in the right column, overlaid with the calculated fundamental oscillation frequencies, $f_i(\nu)$, from the main text.

As expected from the Fermi pocket picture, the QOs scale with applied field as $\frac{\phi_0}{4B_a}$, demonstrated by plotting the frequency axis in units of $\frac{\phi_0}{4B_a}$. Notably, we observe oscillations at applied fields down to $B_a = 116$ mT. At lower fields, only the lowest frequency oscillations are observed. This is explained by the fact that lower frequencies represent FPs with smaller DOS. Therefore, the LLs in these pockets have relatively large energy spacing $E_g$. Since $m_z$ scales with $E_g$ (see below) and due to broadening of the LLs, only the low-DOS FPs have large enough $E_g$ to prevent overlapping of the broadened LLs at low $B_a$. As $B_a$ is increased, higher frequency oscillations (large-DOS pockets) are discernable and $B_z^{ac}$ grows, since all gap sizes are increased.

### Displacement field dependence of quantum oscillations

The measurements of QOs vs. $\nu$ in Figs. 2,3 were performed by sweeping both the backgate voltage $V_{bg}^{dc}$ and the topgate voltage $V_{tg}^{dc}$, which varies both the carrier concentration $n$ and the effective transverse displacement field $D = \varepsilon(V_{tg}^{dc} - V_{bg}^{dc})/2d$, where $\varepsilon = 3.76$ is the dielectric constant of hBN and $d \approx 30$ nm is the thickness of top and bottom hBN. Recent studies have shown a strong effect of $D$ on the electronic structure near CNP in BLG without a moiré (*24, 59–61*). Our transport measurements on BLG/hBN show similar effects near CNP, in particular enhancement of the gap at CNP with $|D|$. In contrast, no changes with $D$ are visible in transport measurements at high doping. Figure S3 shows the QOs measured at four different values of $D$, in the high hole-doping region $\nu \lesssim -3.5$ in Device 3. We find that the displacement field has essentially no effect on QOs within the accessible range of parameters in accord with the transport data. These results are consistent with our BS calculations that show enhancement of the gap at CNP with $|D|$, but no essential effect on the structure of the bands at higher energies.

**Quantum oscillations in Device 2**

In addition to Device 1 with hBN twist angle of $\theta \cong 0.70°$ presented in the main text, we have measured the local QOs in Device 2 with $\theta \cong 1.19°$ and in Device 3 with $\theta \cong 1.37°$. Figure S4 shows QOs in Device 2, similar to main text Fig. 3. In this device the experimentally tunable doping range is smaller. At low $\nu$ periodic QOs are discernable in Fig. S4C, corresponding to four-fold degenerate LLs with $\nu$ independent frequency over the entire region $|\nu| \leq 3.5$, as seen in the FFT in Fig. S4E. For $|\nu| \geq 3.5$, the FFT is similar to Device 1, reflecting the fact that the BS and the FPs are not greatly impacted by the small change in hBN twist angle.

**Band structure calculations**

The continuum model (*41*) is used to calculate the band structure of Bernal BLG on hBN. The lattice vectors of BLG are set as

$$\boldsymbol{a}_1 = a_0\left(\frac{1}{2}, \frac{\sqrt{3}}{2}\right), \ \boldsymbol{a}_2 = a_0\left(-\frac{1}{2}, \frac{\sqrt{3}}{2}\right),$$

where $a_0 = 0.246$ nm is the graphene lattice constant. The sublattices $A$ and $B$ are located at $\boldsymbol{\delta}_A = (0,0)$ and $\boldsymbol{\delta}_B = \frac{a_0}{\sqrt{3}}(0,1)$. The corresponding reciprocal lattice vectors are

$$\boldsymbol{b}_1 = \frac{4\pi}{\sqrt{3}a_0}\left(\frac{\sqrt{3}}{2}, \frac{1}{2}\right), \ \boldsymbol{b}_2 = \frac{4\pi}{\sqrt{3}a_0}\left(-\frac{\sqrt{3}}{2}, \frac{1}{2}\right).$$

hBN has a similar crystal structure, with a larger lattice constant $a_{BN} = 0.2504$ nm $\cong 1.018a_0$. The mismatch between the BLG and hBN lattices and the twist angle $\theta$ give rise to a moiré pattern. The superlattice vectors and reciprocal lattice vectors are given by

$$\boldsymbol{L}_i^M = (\boldsymbol{1} - R^{-1}M^{-1})^{-1}\boldsymbol{a}_i \ (i = 1,2)$$

$$\boldsymbol{G}_i^M = (\boldsymbol{1} - M^{-1}R)\boldsymbol{b}_i \ (i = 1,2)$$

where $R$ is the rotation matrix by $\theta$ and $M = (1 + \varepsilon)\boldsymbol{I}$ is the isotropic lattice mismatch by a factor $(1 + \varepsilon) = 1.018$.

The BLG is coupled to hBN by the interlayer electron hopping. The Hamiltonian is given by

$$H = H_{bi} + H_{BN} + H_{tb},$$

where $H_{bi}$ and $H_{BN}$ are BLG and hBN Hamiltonians, given by

$$H_{bi} = \sum_{\mathbf{q},\zeta,s,\tau_1,\tau_2} a_{s,\zeta}^\dagger(q)h_{bi,\zeta,s}(q)a_{s,\zeta}(q),$$

$$H_{BN} = \sum_{\mathbf{q},\zeta,s} b_{s,\zeta}^\dagger(q)h_{BN,\zeta,s}(q)b_{s,\zeta}(q).$$

Here $a_{s,\zeta}^\dagger(q) = \{a_{s,\zeta,A_1}^\dagger(q), a_{s,\zeta,B_1}^\dagger(q), a_{s,\zeta,A_2}^\dagger(q), a_{s,\zeta,B_2}^\dagger(q)\}$ is the creation operator for an electron with spin index $s = \uparrow, \downarrow$ and valley index $\zeta = \pm 1$ in the sublattices basis $\{A_1, B_1, A_2, B_2\}$ for the BLG, and $b_{s,\zeta}^\dagger(q) = \{b_{s,\zeta,A_1}^\dagger(q), b_{s,\zeta,B_1}^\dagger(q)\}$ is the creation operator for an electron in the sublattices basis $\{A_1, B_1\}$ for the hBN. The momentum $\boldsymbol{q}$ is defined relative to the $K_\zeta$ valley. $h_{bi,\zeta,s}(q)$ and $h_{BN,\zeta,s}(q)$ are given by

$$h_{bi,\zeta,s}(q) = \begin{pmatrix} 0 & v\hat{\pi}^\dagger & -v_4\hat{\pi}^\dagger & v_3\hat{\pi}^\dagger \\ v\hat{\pi} & 0 & \gamma_1 & -v_4\hat{\pi}^\dagger \\ -v_4\hat{\pi} & \gamma_1 & 0 & v\hat{\pi}^\dagger \\ v_3\hat{\pi} & -v_4\hat{\pi} & v\hat{\pi} & 0 \end{pmatrix},$$

$$h_{BN,\zeta,s}(q) \approx \begin{pmatrix} V_N & 0 \\ 0 & V_B \end{pmatrix}.$$

In all the calculations the mini Brillouin zone is sampled with a grid spacing of $10^{-4} nm^{-1}$ in reciprocal space and the dispersion of hBN is ignored.

Similar to the Bistritzer-MacDonald continuum model for twisted bilayer graphene (*62*), the BLG and hBN interlayer coupling is given by

$$H_{tb} = \sum_{\boldsymbol{q},\zeta,s} a_{s,\zeta,\{A_1,B_1\}}^\dagger(\boldsymbol{q}) \left( T_{\boldsymbol{q}_b,\zeta}(\boldsymbol{q},\boldsymbol{q}') + T_{\boldsymbol{q}_{tr},\zeta}(\boldsymbol{q},\boldsymbol{q}') + T_{\boldsymbol{q}_{tl},\zeta}(\boldsymbol{q},\boldsymbol{q}') \right) b_{s,\zeta,\{A_1,B_1\}}(\boldsymbol{q}').$$

Three interlayer hopping processes couple $\boldsymbol{q}$ and $\boldsymbol{q}'$ with the momentum transfer $\boldsymbol{q} - \boldsymbol{q}' = \{\boldsymbol{q}_b, \boldsymbol{q}_{tr}, \boldsymbol{q}_{tl}\}$, where $\boldsymbol{q}_b = -(2G_1^M + G_2^M)/3$, $\boldsymbol{q}_{tr} = (G_1^M + 2G_2^M)/3$, $\boldsymbol{q}_{tl} = (G_1^M - G_2^M)/3$. $G_1^M$ and $G_2^M$ are the reciprocall lattice vectors of the mini-Brillouin zone. The interlayer hopping matrix is given by

$$T_{\boldsymbol{q}_b,\zeta}(\boldsymbol{q},\boldsymbol{q}') = \begin{pmatrix} t_{BC} & 0 \\ 0 & t_{NC} \end{pmatrix} \begin{pmatrix} w & 1 \\ 1 & w \end{pmatrix} \delta_{\boldsymbol{q}-\boldsymbol{q}',\boldsymbol{q}_b}$$

$$T_{\boldsymbol{q}_{tr},\zeta}(\boldsymbol{q},\boldsymbol{q}') = \begin{pmatrix} t_{BC} & 0 \\ 0 & t_{NC} \end{pmatrix} \begin{pmatrix} w & e^{-i\zeta\frac{2\pi}{3}} \\ e^{i\zeta\frac{2\pi}{3}} & w \end{pmatrix} \delta_{\boldsymbol{q}-\boldsymbol{q}',\boldsymbol{q}_{tr}}$$

$$T_{\boldsymbol{q}_{tl},\zeta}(\boldsymbol{q},\boldsymbol{q}') = \begin{pmatrix} t_{BC} & 0 \\ 0 & t_{NC} \end{pmatrix} \begin{pmatrix} w & e^{i\zeta\frac{2\pi}{3}} \\ e^{-i\zeta\frac{2\pi}{3}} & w \end{pmatrix} \delta_{\boldsymbol{q}-\boldsymbol{q}',\boldsymbol{q}_{tl}}$$

where $t_{BC}$ is the tunneling between boron and carbon, $t_{NC}$ is the tunneling between nitrogen and carbon, $w = \frac{t_{AA}}{t_{AB}}$ is the phenomenological modeling of lattice relaxation, and $t_{AA}$ and $t_{AB}$ are the tunneling amplitudes in *AA* and *AB* regions, respectively. The real-space moiré unit cell has different regions of local stacking orientation. There are *AA* regions, where both carbon atoms are directly above the hBN atoms and *AB* regions, where only one carbon atom is directly above one of the hBN atoms. Due to the large size of the moiré supercell, the lattice has the freedom to relax and minimize the area of the high energy *AA* alignment regions in favor of *AB* alignment, which is modeled using the effective parameter $w$. Additionally, we define $r_{NB} \equiv t_{NC}/t_{BC}$. In previous studies of hBN induced moiré, only a limited range of $r_{NB}$ has been considered, and the effect of $w$ has been ignored.

### Band structure fitting

Upon increasing the carrier density $n$, the FPs are populated according to their ratios of DOS. For a given band structure, we can thus derive the QO frequencies $f_i(n)$ of the different pockets from the calculated DOS and compare with the experimental data. Our high sensitivity to the oscillation frequencies thus allows detailed fitting of the BS and extracting previously unknown BS parameters.

The presented simulations are done for $\theta = 0.75°$, close to the twist angle of Device 1. We find that the common literature values of $\hbar v = 5.27$ eVÅ, $\gamma_1 = 0.30$ eVÅ, $V_N = -1.40$ eVÅ, and $V_B = 3.34$ eV ($63, 64$) provide a good fit to the experimental data. However, the inter-BLG hopping parameters $v_3$ and $v_4$, which are responsible for trigonal warping, have a wide range of experimentally reported values ($64$). We find that the main effect of $v_4$ is lifting the three-fold symmetry and thus splitting the 12-fold degenerate pockets that appear at $\nu \sim \pm 4$ (see Movie S3). Since this is inconsistent with the data, we set $v_4 = 0$. We then test the range of the reported values for $v_3 = 0.21 - 0.67$ eVÅ. We find that $v_3$ affects the relative size of V2 and V3 pockets and find an optimal value of $v_3 = 0.51$ eVÅ.

The tunneling strength between the hBN substrate and graphene has not been studied experimentally in detail to date, and therefore we explored the full phase space of its parameters. Figure S5 shows the comparison of parameters used for BS calculations in Ref. ($41$) with $r_{NB} = 1$ and $t_{BC} = 0.152$ eV (left column) and Ref. ($45$) with $r_{NB} = 0.67$ and $t_{BC} = 0.144$ eV (middle column), both of which ignored the effect of lattice relaxation ($w = 1$). Neither of these parameter sets match experiment well.

We find that the magnitude of $t_{BC}$ does not strongly affect the calculated QOs and therefore set $t_{BC} = 0.144$ eV while varying $r_{NB}$ and $w$. Without lattice relaxation, $w = 1$, we find a good fit only for $r_{NB} < 0.2$. However, such a small value of $r_{NB}$ has no physical justification. Allowing for lattice relaxation, we find optimal values to fit experimental data to be $r_{NB} = 0.5, w = 0.5$ (Figs. S5C,F). These parameters are used in the main text.

Note that as the tunneling strength between hBN and graphene is reduced, the gaps between the remote bands decrease, reducing the magnetic breakdown field. The BS with parameters from Ref. ($41$) has a full gap at $\nu = -4$ (Fig. S5A), whereas the BS with our derived parameters only has a slight minimum in DOS (Fig. S5C), in good agreement with the small peak in $\rho_{xx}$ (Fig. 1C). This weak coupling between hBN and graphene causes highly overlapping bands with parallel ridges separated by small gaps over extended ranges of energy, as seen in Fig. 5B, thereby leading to an extremely low $B_{MB}$.

**Description of quantum oscillations**

It is instructive to describe the dHvA QOs in 2D systems in terms of competition between the diamagnetic and paramagnetic contributions to orbital magnetization. Semiclassically, an electron in a parabolic band with effective mass $m^*$ and energy $|\epsilon| = \hbar^2 k^2 / 2m^*$ moves in a cyclotron orbit with a radius $R = \frac{\hbar k}{e B_a}$ at cyclotron frequency $\omega_c = e B_a / m^*$. This cyclotron motion gives rise to a diamagnetic moment $\mu_e$ determined by the current $I = -e \omega_c / 2\pi$ times the cyclotron area $A = \pi R^2$, $\mu_e = -e \omega_c R^2 / 2 = -|\epsilon| / B_a$. In the case of sharp LLs, only discrete energy levels are allowed, $E_j = \left(j + \frac{1}{2}\right) \hbar \omega_c$, where $j$ is the LL index. An electron added to the highest partially filled compressible LL thus contributes a diamagnetic moment, $\mu_e = m_z = \frac{dM_z}{dn} = -|E_j| / B_a$. Equivalently, by increasing $B_a$ at constant carrier density, the chemical potential $\mu$ of the compressible LL increases, giving rise to $m_z = -\frac{d\mu}{dB_a}\Big|_n = -\frac{dE_j}{dB_a} = -E_j / B_a$ leading to an identical result. Filling one complete LL by carriers, thus contributes a total magnetization of $M_{dia} = n_{eLL} \mu_e = -e E_j / h = -\left(j + \frac{1}{2}\right) e \omega_c / 2\pi$ per unit area, where $n_{eLL} = e N B_a / h$ is the density of a full LL and $N$ is the LL degeneracy which we take here to be 1 for simplicity. This Landau diamagnetism arises from equilibrium currents flowing in the compressible bulk states provided the LLs are very sharp, namely the chemical potential does not change during the process of filling the compressible states with $\mu = E_j$. In the incompressible gaped states, in contrast, the response is paramagnetic. Upon crossing the gap, the chemical potential increases abruptly by $\Delta \mu = \Delta E_j = \hbar \omega_c$, giving rise to an increase in the total free energy of $n\Delta\mu$ and hence a jump in the total magnetization of $M_{para} = \frac{n\Delta\mu}{B_a} = \frac{(j+1)e\omega_c}{2\pi}$. Note that $M_{para} \cong -M_{dia}$ for large $j$. The paramagnetic contribution arises due to topological currents in the incompressible edge states that counterflow the diamagnetic currents in the compressible edge states (65). The magnetization oscillations with carrier density attain a maximal value when the LLs are sharp, with diamagnetic currents generated in the compressible state while $\mu$ is maintained constant, followed sequentially by paramagnetic currents generated in the incompressible state while $n$ remains constant. In the presence of LL broadening, $\mu$ increases while filling the compressible LLs and $n$ grows also in the incompressible states resulting in concurrent rather than sequential generation of diamagnetic and paramagnetic currents, partially cancelling each other and decreasing the amplitude of the magnetization oscillations.

## Frequency of quantum oscillations

The magnetization oscillations at the fundamental frequency can be approximated at low temperatures by $M_z = M_0 \sin\left(2\pi \frac{\phi_0}{N} \frac{n}{B_a}\right)$, where $\phi_0 = h/e$ is the flux quantum and we have suppressed for simplicity the constant phase $\varphi$ which equals $-\pi$ for massive particles and 0 for Dirac fermions. The frequency of the QOs vs. $n$ is thus given by $f_n = \frac{\phi_0}{N} \frac{1}{B_a}$, while the frequency vs. $1/B_a$ is $f_B = \frac{\phi_0}{N} n$. In the presence of multiple Fermi surfaces, both the frequency and the amplitude of the QO are altered. For concreteness, we inspect two parabolic bands, $\epsilon_1 = \frac{2\pi\hbar^2 n_1}{m_1^*}$ and $\epsilon_2 = \frac{2\pi\hbar^2 n_2}{m_2^*} + \epsilon_0$ shifted in energy by $\epsilon_0 > 0$. In thermal equilibrium, $\epsilon = \epsilon_1 = \epsilon_2$ and $n = n_1 + n_2$, with $n_1 = n$ and $n_2 = 0$ for $n \leq n_0 = \epsilon_0 m_1^*/2\pi\hbar^2$, and $n_1 = n_0 + \frac{(n-n_0)m_1^*}{(m_1^*+m_2^*)}$ and $n_2 = \frac{(n-n_0)m_2^*}{(m_1^*+m_2^*)}$ for $n > n_0$. When both bands are occupied, the magnetization oscillations display two fundamental frequencies, $M_z = M_1 \sin\left(2\pi \frac{\phi_0}{N} \frac{n_1}{B_a}\right) + M_2 \sin\left(2\pi \frac{\phi_0}{N} \frac{n_2}{B_a}\right)$.

Traditionally, the QOs are measured vs. $B_a$, in which case the ratio of the two frequencies is given by the ratio of the Fermi surface areas (20, 22), $\frac{f_{B1}}{f_{B2}} = \frac{n_1}{n_2} = \frac{S_1}{S_2}$. This conjecture, however, does not hold for QO measured vs. carrier density $n$. By using the expressions for $n_1$ and $n_2$, one attains $M_z = M_1 \sin(2\pi f_{n1}n + 2\pi f_{n2}n_0) + M_2 \sin(2\pi f_{n2}n - 2\pi f_{n2}n_0)$, where $f_{n1} = \frac{m_1^*}{(m_1^*+m_2^*)} \frac{\phi_0}{NB_a}$ and $f_{n2} = \frac{m_2^*}{(m_1^*+m_2^*)} \frac{\phi_0}{NB_a}$. Since the arguments proportional to $n_0$ are constant phases, the ratio of the frequencies vs. $n$ is thus given by the ratio of the effective masses or of the density of states $\mathcal{N}(\epsilon)$, $\frac{f_{n1}}{f_{n2}} = \frac{m_1^*}{m_2^*} = \frac{\mathcal{N}_1(\epsilon)}{\mathcal{N}_2(\epsilon)}$, rather than by the ratio of the Fermi surfaces. This difference can be understood as follows. The term $2\pi f_{n2}n_0$ originates from the LLs residing at $\epsilon \leq \epsilon_0$ in band 1 which are stationary for a given $B_a$ and just add a constant phase to the oscillations. These lower LLs do not affect the process of filling of the higher LLs upon increasing $n$, and hence the ratio of the occupations and of the oscillation frequencies of the two bands for $\epsilon > \epsilon_0$ is given by $\frac{\mathcal{N}_1(\epsilon)}{\mathcal{N}_2(\epsilon)}$. When varying the field, in contrast, the lower LLs shift with $B_a$ and hence the phase $2\pi f_{n2}n_0$ grows with $1/B_a$, contributing to the QOs of the higher LLs with the ratio of the two frequencies determined by the ratio of the Fermi surfaces $\frac{S_1(\epsilon)}{S_2(\epsilon)}$. Note that in both cases the sum of the frequencies equals the original frequency of a single band, $f_{B1} + f_{B2} = f_B = \frac{\phi_0}{N} n$ and $f_{n1} + f_{n2} = f_n = \frac{\phi_0}{N} \frac{1}{B_a}$.

Using the above description, we can derive the fundamental frequencies of the QOs expected from the band structure of Fig. 4A. For $|\nu| \lesssim 3.5$ the flat bands V1 and C1 have a single fourfold degenerate Fermi surface as shown in Fig. 4B, and hence $f_n = \frac{\phi_0}{4B_a}$. For higher doping, the trigonal warping and the overlapping bands give rise to multiple FPs and consequently several oscillation frequencies.

**Amplitude of magnetization oscillations**

As pointed out originally by Peierls (*66*), the maximal peak-to-peak amplitude of the magnetization oscillations normalized by the total carrier density is $\frac{\Delta M}{n} = \frac{\hbar \omega_c}{B_a} = 2\mu_B^*$, where $\mu_B^* = \frac{e\hbar}{2m^*} = \frac{m_e}{m^*}\mu_B$ is the effective Bohr magneton. For 2DEG in GaAs, for example, $\mu_B^* \cong 15\mu_B$, as has been demonstrated experimentally (*13–16, 21, 67, 68*). Rather than measuring the magnetization averaged over all the carriers, our experiment determines the differential magnetization $m_z$ at the Fermi energy $\epsilon_F$. A carrier added to the system in a fully compressible state carries a magnetic moment of $m_z = \frac{dM_z}{dn} = -\frac{|\mu|}{B_a}$. Our measured value of $m_z \cong -500\mu_B$ at $B_a = 300$ mT thus corresponds to $|\mu| \cong 9$ meV. This is a lower bound on $|\mu|$, since as discussed above, broadening of the LLs and the presence of additional bands reduces $m_z$ and the estimated value of $|\mu|$. The Fermi pockets in Fig. 4C each span a chemical potential range of up to $|\mu| \cong 100$ meV consistent with the experimental lower bound. Note that the low-frequency QOs correspond to pockets with lower DOS and thus larger LL energy gaps. As a result, they are more robust against LL broadening and show more pronounced magnetization oscillations as seen in Fig. 3E.

Note also that for a given oscillation amplitude of the total magnetization $M_z = M_0 \sin(2\pi f_n n)$, the measured differential $m_z = \frac{dM_z}{dn} = 2\pi f_n M_0 \sin(2\pi f_n n)$ is proportional to the oscillation frequency $f_n$. As a result, $m_z$ vanishes for $f_n$ approaching zero, explaining the suppressed intensity of the experimental fundamental frequency lines at very low frequencies in Fig. 4D.

**Position dependence of local quantum oscillations**

Local QOs are a highly sensitive probe of the local band structure. Figure 3A shows oscillations in $B_z^{ac}$ as a function of $\nu$ and position across the sample with spatial resolution of our SOT of about 150 nm. Variations in the local BS may result from twist angle disorder between BLG and hBN, leading to a position dependence of the QOs. We estimate the twist angle variations of about ~3% across the sample with a characteristic length scale of about a μm, substantially smaller than the twist angle disorder reported for magic-angle graphene (*69*). The variations in the local QOs are best appreciated by inspecting Movie S2, which shows FFT as a function of position across the sample. While there is only a slight spatial variation in the values of the fundamental frequencies, strong variations in the amplitudes of the QOs are found. For example, the amplitudes of $f_{V3X}$ and $f_{V3Y}$ (along the purple lines) are highly position dependent, showing modulation in their relative intensity. This can be explained by shifts in the relative phase between the two close oscillation frequencies giving rise to the observed beating. Local strain may induce such phase shifts through strain-induced pseudomagnetic fields. Another possible source of intensity variation is a position-dependent lifetime broadening of the LLs governed by interpocket scattering. Notably, the intensity of the CMB oscillations (along light green lines) varies significantly with position. This may arise from position dependence of the lattice relaxation parameter $w$ that governs the gap size $\Delta k$ between the Fermi pockets. Such $\Delta k$ disorder affects both the lifetime broadening of the fundamental frequencies and the intensity of the CMB QOs. Importantly, at our $B_a = 0.3$ T, the characteristic cyclotron radius of both the fundamental and the CMB orbits is $R_c = \frac{\phi_0}{2\pi B_a}\left(\frac{S}{\pi}\right)^{1/2} \cong 500$ nm, where $S \cong S_{mBZ}/2$ and $S_{mBZ} = 0.36$ nm$^{-2}$. Since the real-space size of the QO orbits is comparable to the characteristic length scale of BS variations, further theoretical and experimental studies are required for understanding the QOs in the presence of BS disorder.

**Full magnetic breakdown at elevated fields**

At low magnetic fields the tunneling probability $P$ between the FS pockets is negligible. At such fields the carriers can orbit only along the closed FSs of individual pockets (black arrows in Fig. S6), resulting in QOs at fundamental frequencies. At intermediate $B_a$ the finite tunneling probability due to CMB, $0 < P < 1$, results in a network of electron orbits (*25, 27, 70, 71*), giving rise to a multitude of QO frequencies. Figure 5 shows examples of the shortest CMB orbits, which are most pronounced experimentally. With increasing $B_a$, the tunneling probability approaches $P = 1$, leading to a full magnetic breakdown with only one allowable orbit, as depicted by the green trajectory in Fig. S6, which persists over a wide range of carrier densities. As a result, the QOs will display a single frequency given by $f_n = \frac{\phi_0}{4}\frac{1}{B_a}$ vs. $n$ and $f_B = \phi_0 \frac{S}{4\pi^2}$ vs. $1/B_a$, where $S$ is the $k$-space orbit area. In Fig. S6, this area, given by $S = 4\pi^2(n/4)$, is larger than the mBz area, $S > S_{mBz} = 4\pi^2 n_0$, corresponding to carrier density $n > 4n_0$ determined by the back-gate voltage. Thus the full-magnetic-breakdown orbit follows essentially the FS (green trajectory) corresponding to the original BLG band structure in the absence of moiré potential at the same density $n$. As a result, the information about the moiré minibands is mostly lost as evidenced by the disappearance of the van Hove singularity and $n_H$ sign reversals (Fig. 1D, blue curve). Simultaneously, the Landau fan restores its default form with four-fold LLs originating from CNP. This explains the reoccurrence of LLs at elevated fields that extrapolate to CNP for $|\nu| > 4$ as observed in transport measurements shown in Figs. S1A,B.

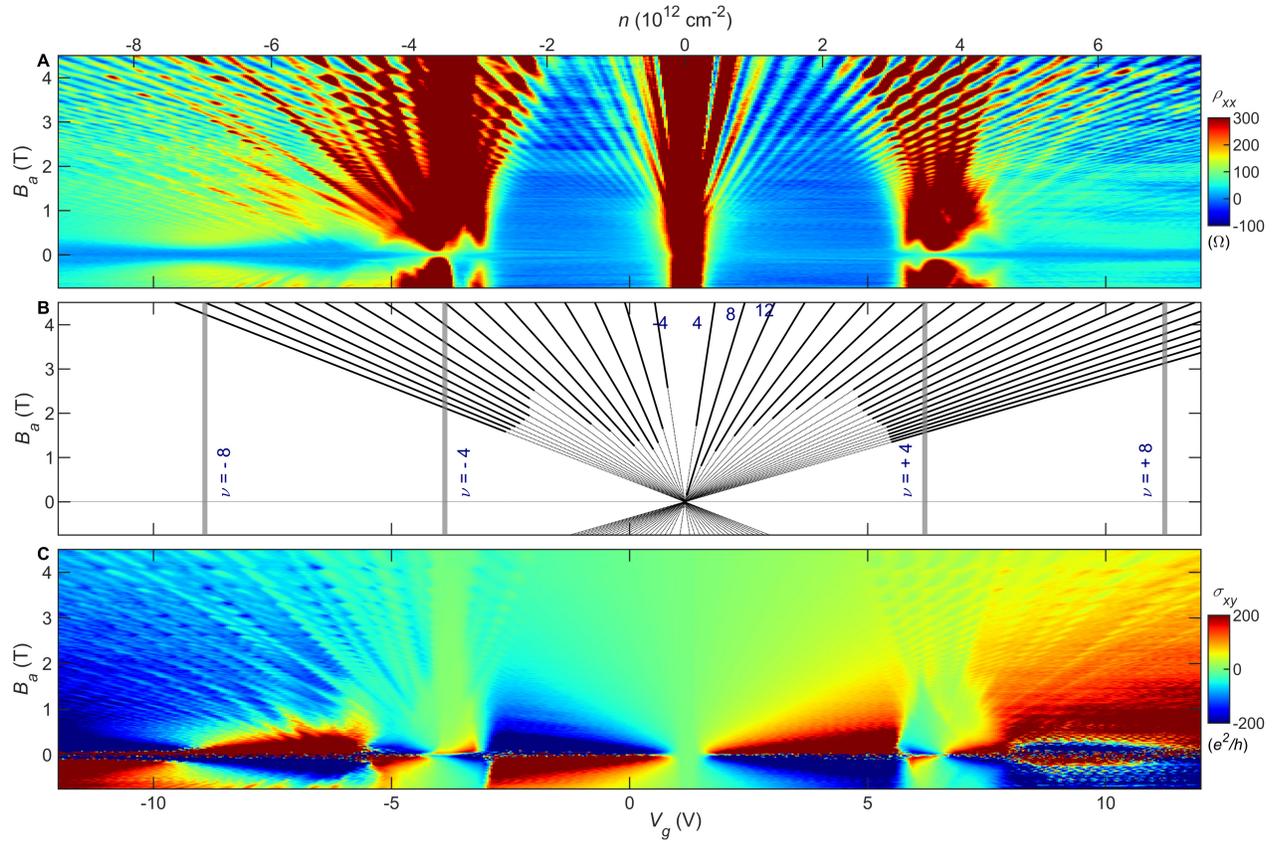

**Fig. S1. Transport measurements and Landau fan.** (**A**) Longitudinal resistivity $\rho_{xx}$ for Device 1 at $T = 300$ mK as in Fig. 1A over a reduced range of $n$. (**B**) Fits to the Landau fan originating from CNP. The four-fold degenerate Landau levels are indicated by thick diagonal lines along with their corresponding Chern numbers, while the thin lines are linear extrapolations. Specific fillings $\nu$ of the moiré supercell are indicated by vertical grey lines. (**C**) Hall conductivity $\sigma_{xy}$. High values of $\rho_{xx}$ and $\sigma_{xy}$ are saturated for clarity.

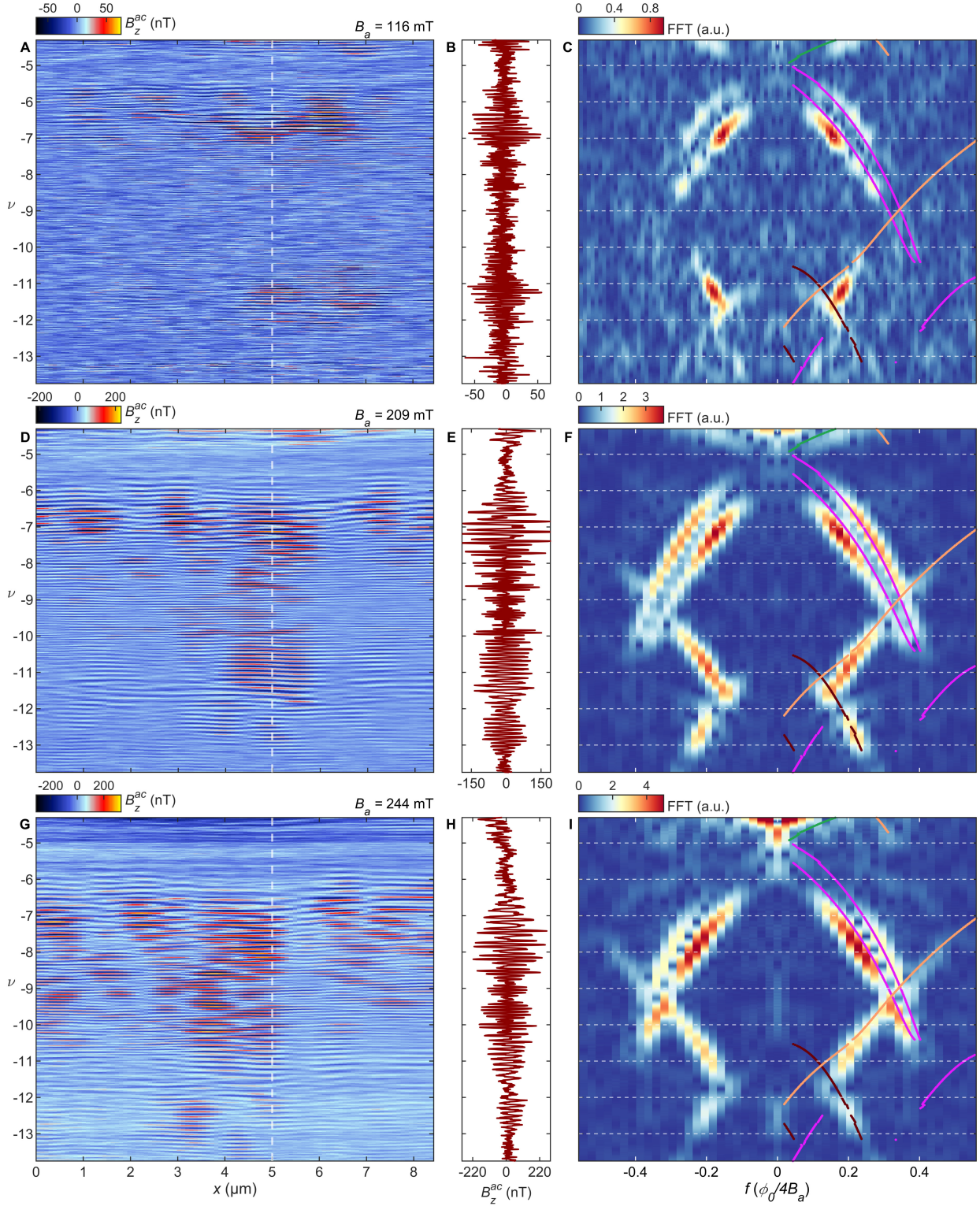

**Fig. S2. Magnetic field dependence of quantum oscillations.** (**A**) QOs in Device 1 vs. $\nu$ at $B_a = 116$ mT with $V_g^{ac} = 10$ mV. (**B**) Line cut along the white dotted line in (A). (**C**) FFT of the data in (B) using a narrow window of $\delta\nu = 1.11$ around a given $\nu$, overlaid with the calculated fundamental QOs frequencies from the main text. (**D-I**) Same as (A-C) with $V_g^{ac} = 20$ mV at $B_a = 209$ mT and 244 mT. QOs over the full range of $\nu$ are shown in Fig. 3 at $B_a = 300$ mT with $V_g^{ac} = 10$ mV.

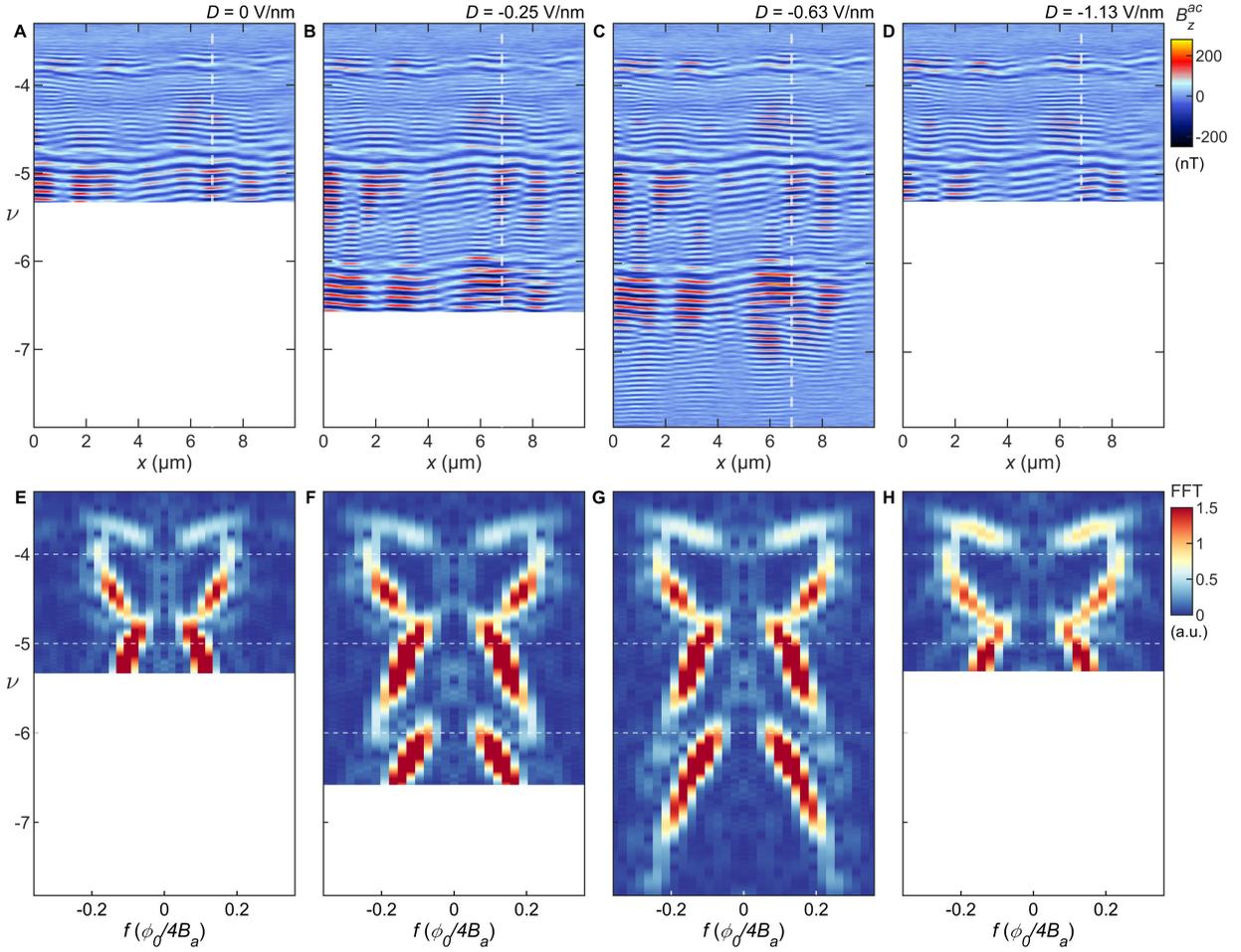

**Fig. S3. Displacement field dependence of quantum oscillations.** (**A-D**) QOs in Device 3 at $B_a = 209$ mT for $D = 0$ V/nm (A), $-0.25$ V/nm (B), $-0.63$ V/nm (C), and $-1.13$ V/nm (D) using $V_{bg}^{ac} = 50$ mV. (**E-H**) Corresponding FFT along the white dashed lines in (A-D) using a narrow window of $\delta\nu = 0.44$ around a given $\nu$. The displacement field has weak effect on QOs apparently due to its effective screening at high carrier doping of BLG.

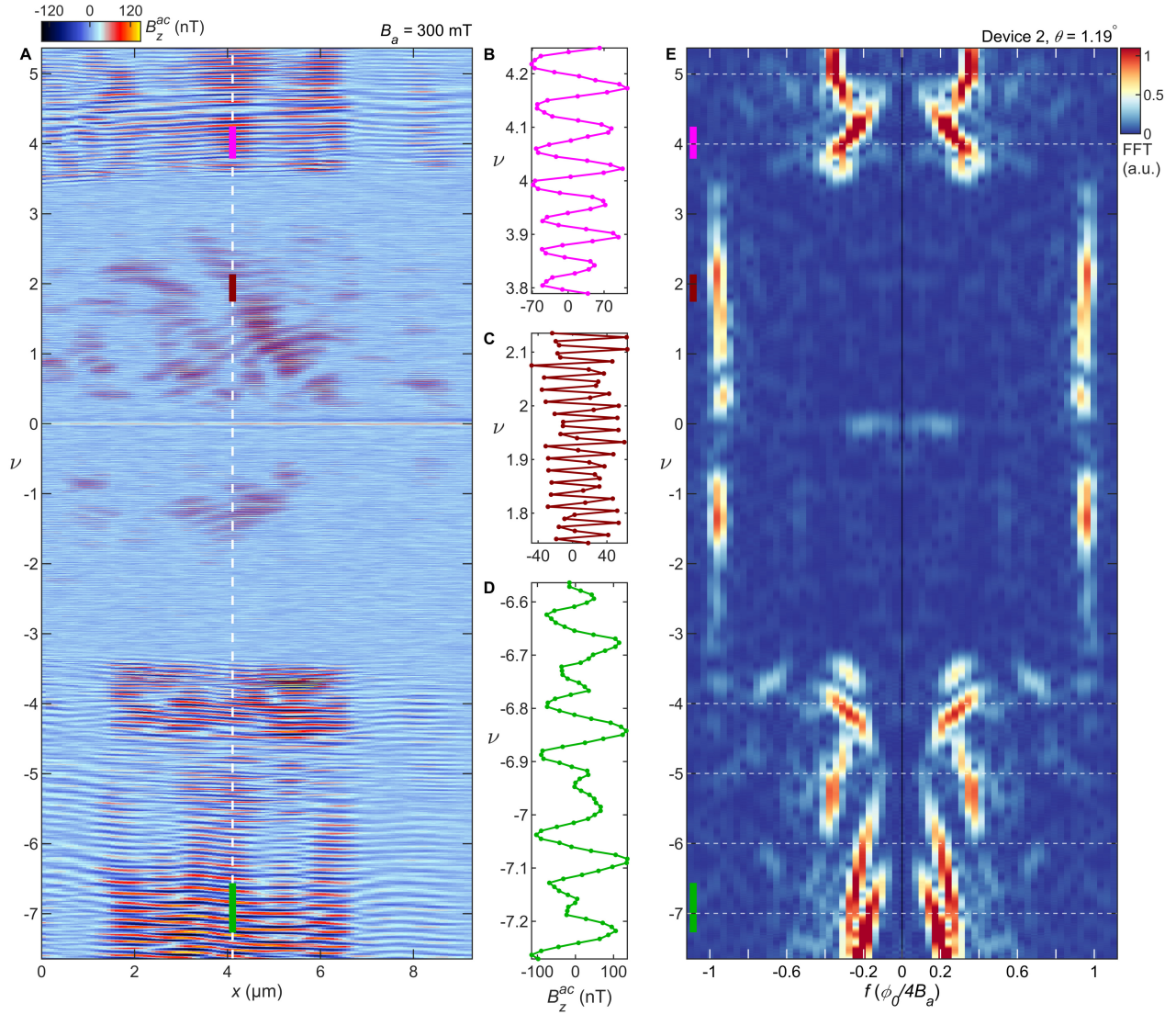

**Fig. S4. Quantum oscillations in Device 2.** (**A-E**) Same as Fig. 3 but for Device 2. The tunable range of $\nu$ is smaller in this device as compared to Device 1 from the main text.

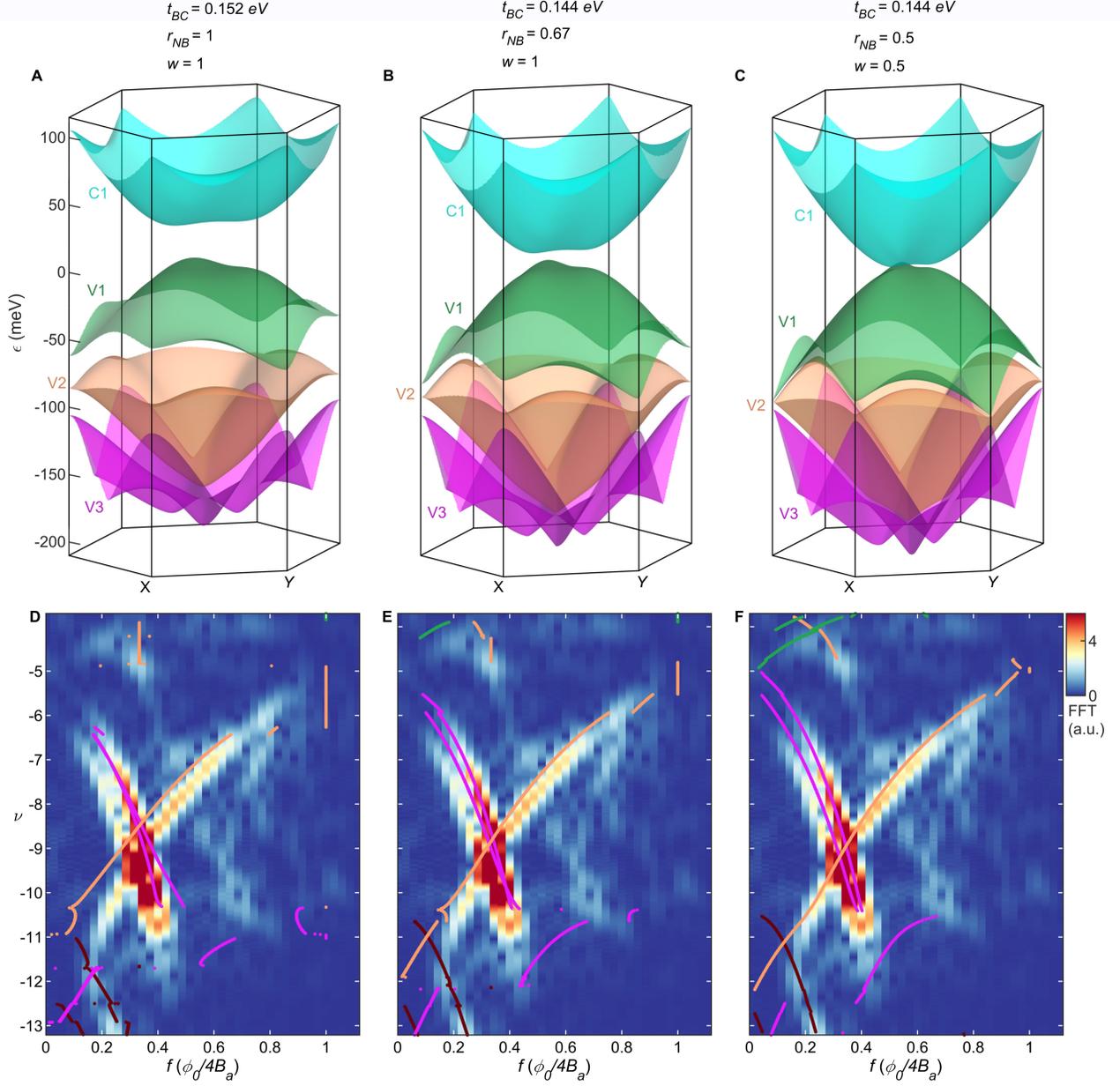

**Fig. S5. Fitting band structure parameters.** (**A**) Calculated band structure of BLG aligned to hBN with $\theta = 0.75°$ with hBN tunneling parameters taken from Ref. (*41*) $t_{BC} = 0.152\ eV$, $r_{NB} = 1$, and $w = 1$. (**B**) Same as (A) with hBN tunneling parameters taken from Ref. (*45*) with $t_{BC} = 0.144\ eV$, $r_{NB} = 0.67$, and $w = 1$. (**C**) Same as (A) with our derived hBN tunneling parameters taken to best fit the experimental data: $t_{BC} = 0.144\ eV$, $r_{NB} = 0.5$, and $w = 0.5$. (**D-F**) Experimental FFT from main text overlaid with the calculated fundamental QO frequencies for band structures in (A-C).

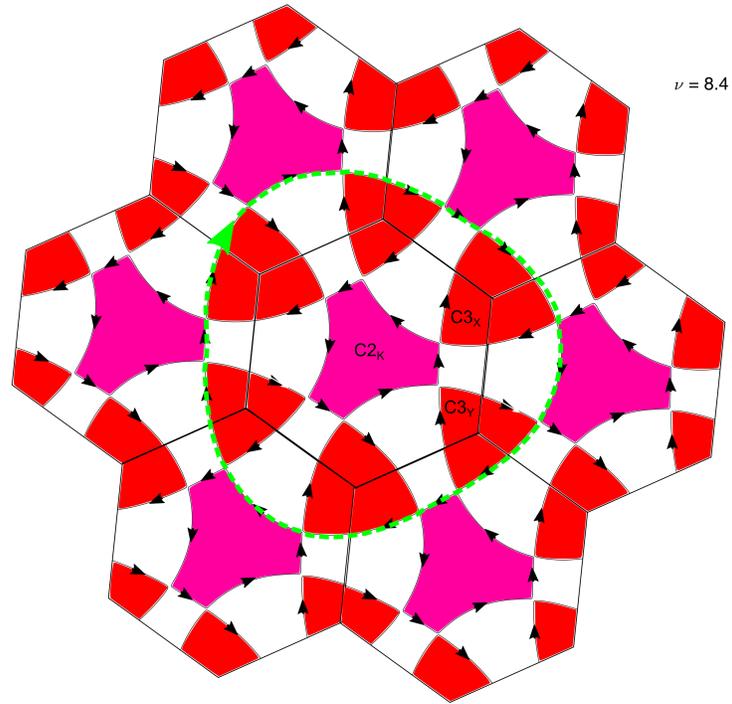

**Fig. S6. Full magnetic breakdown.** Constant energy band structure cut at $\nu = 8.4$ over an extended Brillouin zone. The dashed green trajectory indicates the full magnetic breakdown orbit that traces the original BLG Fermi surface with vanishing moiré potential.

**Movie S1. Tomographic rendering of the quantum oscillations in the local magnetization**. Evolution of $m_z$ as a function of the spatial axes $x$ and $y$ and the filling factor $\nu$, at $B_a = 334$ mT in Device 1. Different two-dimensional slices in the $x - \nu$, $y - \nu$, and $x - y$ planes are presented sequentially. The movie shows quantum oscillations of positive and negative $m_z$ with values in excess of 500 $\mu_\mathrm{B}$/electron, evolving across the sample upon varying $\nu$.

**Movie S2. Position dependence of the local quantum oscillations**. Left panel: $B_z^{ac}(x, \nu)$ measured in Device 1 along the dotted black line in Fig. 2A at $B_a = 300$ mT, reproduced from Fig. 3A. Middle panel: line cut of $B_z^{ac}(\nu)$ along the white dotted line in the left panel. Right panel: FFT of middle panel performed over a narrow window of $\delta\nu = 1.11$ around $\nu$. The FFT frequency is in units of $\phi_0/4B_a$. At $|\nu| \lesssim 3.5$ the quantum oscillations occur at $f = \frac{1}{4}\frac{\phi_0}{B_a}$ independent of the location $x$ and the filling factor $\nu$, arising from conventional four-fold degenerate QHE LLs. For $|\nu| \gtrsim 3.5$ the low-frequency oscillations are governed by multiple Fermi surfaces and magnetic breakdown as indicated by the color-coded solid lines.

**Movie S3. Calculated BLG moiré band structure**. Constant-energy band structure cuts showing the evolution of the closed Fermi surface electron and hole pockets with $\epsilon$ and $\nu$. For $|\nu| \gtrsim 3.5$ multiple Fermi pockets are formed in the different overlapping moiré minibands (Fig. 4A). The FPs are separated by small momentum gaps at the touching points over the entire range of $\nu$. The legend on the right indicates the bands, the areas of the FPs, and their relative sizes.